\documentclass[sigconf]{acmart}
\usepackage{tabularx}
\usepackage{multirow}
\usepackage{csvsimple}
\usepackage{mhchem}
\usepackage{graphicx}
\usepackage{subcaption}
\usepackage{mwe}
\usepackage{float}
\usepackage{placeins}
\usepackage{amsfonts}
\usepackage{booktabs}
\usepackage{siunitx}
\usepackage{array}
\usepackage{pdfpages}

\def\BibTeX{{\rm B\kern-.05em{\sc i\kern-.025em b}\kern-.08emT\kern-.1667em\lower.7ex\hbox{E}\kern-.125emX}}
    
\copyrightyear{2019} 
\acmYear{2019} 
\setcopyright{acmlicensed}
\acmConference[e-Energy '19]{Proceedings of the Tenth ACM International Conference on Future Energy Systems}{June 25--28, 2019}{Phoenix, AZ, USA}
\acmBooktitle{Proceedings of the Tenth ACM International Conference on Future Energy Systems (e-Energy '19), June 25--28, 2019, Phoenix, AZ, USA}
\acmPrice{15.00}
\acmDOI{10.1145/3307772.3335321}
\acmISBN{978-1-4503-6671-7/19/06}

\newcolumntype{M}[1]{>{\centering\arraybackslash}m{#1}}
\begin{document}

%
\title[ElecSim: Market Model]{ElecSim: Monte-Carlo Open-Source Agent-Based Model to Inform Policy for Long-Term Electricity Planning}

%

\author{Alexander Kell}
\affiliation{%
  \department{School of Computing}
  \institution{Newcastle University}
  \city{Newcastle upon Tyne}
  \country{UK}
}
\email{a.kell2@newcastle.ac.uk}

\author{Matthew Forshaw}
\affiliation{%
  \department{School of Computing}
  \institution{Newcastle University}
  \city{Newcastle upon Tyne}
  \country{UK}
}
\email{matthew.forshaw@newcastle.ac.uk}

\author{A. Stephen McGough}
\affiliation{%
  \department{School of Computing}
  \institution{Newcastle University}
  \city{Newcastle upon Tyne}
  \country{UK}
}
\email{stephen.mcgough@newcastle.ac.uk}
 
%


%
\begin{abstract}

Due to the threat of climate change, a transition from a fossil-fuel based system to one based on zero-carbon is required. However, this is not as simple as instantaneously closing down all fossil fuel energy generation and replacing them with renewable sources -- careful decisions need to be taken to ensure rapid but stable progress. To aid decision makers, we present a new tool, ElecSim, which is an open-sourced agent-based modelling framework used to examine the effect of policy on long-term investment decisions in electricity generation. ElecSim allows non-experts to rapidly prototype new ideas. 

Different techniques to model long-term electricity decisions are reviewed and used to motivate why agent-based models will become an important strategic tool for policy. We motivate why an open-source toolkit is required for long-term electricity planning.

Actual electricity prices are compared with our model and we demonstrate that the use of a Monte-Carlo simulation in the system improves performance by $52.5\%$. Further, using ElecSim we demonstrate the effect of a carbon tax to encourage a low-carbon electricity supply. We show how a \textsterling40 ($\$50$) per tonne of \ce{CO2} emitted would lead to 70\% renewable electricity by 2050. \vphantom{{\color{red}An interesting note, however, is that starting with a low carbon tax and slowly increasing this by the year 2050 provides similar benefits to a lower, but consistent tax in the long run, due to the high capital costs and long operating periods of generators. This has the benefits of reducing costs as well as providing certainty to investors.}}

\end{abstract}

%
%



%


%

%
\maketitle

\section{Introduction}

The world faces significant challenges from climate change \cite{Masson-Delmotte2018}. A rise in carbon emissions increases the risk of severe impacts on the world such as rising sea levels, heat waves and tropical cyclones \cite{Masson-Delmotte2018}. A survey \cite{Cook2013} showed that 97\% of scientific literature concurs that the recent change in climate is anthropogenic.

 High carbon emitting electricity generation sources such as coal and natural gas currently produce 65\% of global electricity, whereas low-carbon sources such as wind, solar, hydro and nuclear provide 35\% \cite{BP2018}. Hence, to bring about change and reach carbon-neutrality, a transition in the electricity mix is required.

 Due to the long construction times, operating periods and high costs of power plants, investment decisions can have long term impacts on future electricity supply \cite{Chappin2017}. Governments and society, therefore, have a role in ensuring that the negative externalities of emissions are priced into electricity generation. This is most likely to be achieved via carbon tax and regulation to influence electricity market players such as generation companies (GenCos).

Decisions made in an electricity markets may have unintended consequences due to their complexity. A method to test hypothesis before they are implemented would therefore be useful.

Simulation is often used to increase understanding as well as to reduce risk and reduce uncertainty. Simulation allows practitioners to realise a physical system in a virtual model. In this context, a model is defined as an approximation of a system through the use of mathematical formulas and algorithms. Through simulation, it is possible to test a system where real life experimentation would not be practical due to reasons such as prohibitively high costs, time constraints or risk of detrimental impacts. This has the dual benefit of minimising the risk of real decisions in the physical system, as well as allowing practitioners to test less risk-averse strategies.

Agent-based modelling (ABM) is a class of computational simulation models composed of autonomous, interacting agents and model the dynamics of a system. Due to the numerous and diverse actors involved in electricity markets, ABMs have been utilised in this field to address phenomena such as market power \cite{Ringler2016a}. 

This paper presents ElecSim, an open-source ABM that simulates GenCos in a wholesale electricity market. ElecSim models each GenCo as an independent agent and electricity demand. An electricity market facilitates trades between the two. 

GenCos make bids for each of their power plants. Their bids are based on the generator's short run marginal cost (SRMC) \cite{Perloff2012}, which excludes capital and fixed costs. The electricity market accepts bids in cost order, also known as merit-order dispatch. GenCos invest in power plants based on expected profitability.	

ElecSim is designed to provide quantitative advice to policy makers, allowing them to test policy outcomes under different scenarios. They are able to modify a script to realise a scenario of their choice. It can also be used by energy market developers who can test new electricity sources or policy types, enabling the modelling of changing market conditions.



The contribution of this paper is a new open-source framework with example scenarios of varying carbon taxes. We provide curated data, and improve realism via Monte-Carlo sampling. Section \ref{Literature Review} is a literature review. Section \ref{Model} details the model and assumptions made, and Section \ref{Validation and Performance} provides performance metrics and validation. Section \ref{Scenario Testing} details our results. We conclude the work in Section \ref{Conclusion}.

\section{Literature Review}\label{Literature Review}

Live experimentation of physical processes is often not practical. The costs of real life experimentation can be prohibitively high, and can require significant time in order to fully ascertain the long-term trends. There is also a risk that changes can have detrimental impacts and lead to risk-averse behaviour. These factors are true for electricity markets, where decisions can have long term impacts. Simulation, however, can be used for rapidly prototyping ideas. The simulation is parametrised by real world data and phenomena. Through simulation, the user is able to assess the likelihoods of outcomes under certain scenarios and parameters \cite{Law:603360}.

Energy models can typically be classified as top-down macro-economic models or bottom-up techno-economic models~\cite{Bohringer1998}. Top-down models typically focus on behavioural realism with a focus on macro-economic metrics. They are useful for studying economy-wide responses to policies ~\cite{Hall2016}, for example MARKAL-MACRO \cite{Fishbone1981} and LEAP \cite{Heaps2016}. Bottom-up models represent the energy sector in detail, and are written as mathematical programming problems~\cite{Gargiulo2013}. 

It is possible to further categorise bottom-up models into optimisation and simulation models. Optimisation energy models minimise costs or maximise welfare, defined as the material and physical well-being of people ~\cite{Keles2017}. Examples of optimisation models are MARKAL/TIMES~\cite{Fishbone1981} and MESSAGE~\cite{Schrattenholzer1981}. 

However, electricity market liberalisation in many western democracies has changed the framework conditions. Centralised, monopolistic, decision making entities have given way to multiple heterogeneous agents acting for their own best interest~\cite{Most2010}. Policy options must therefore be used to encourage changes to attain a desired outcome. It is proposed that these complex agents are modelled using ABMs due to their non-deterministic nature.


Traditional centralised optimisation models are not designed to  describe a system which is out of equilibrium. Optimisation models assume perfect foresight and risk neutral investments with no regulatory uncertainty. The core dynamics which emerge from equilibrium remain a black-box. For example, the model assumes a target will be reached, and does not provide information for which this is not the case. Reasons for this could be investment cycles which move the model away from equilibrium \cite{Chappin2017}.




\begin{table}[]
	\small	
	\begin{tabular}{M{2cm}M{0.8cm}M{1cm}M{0.8cm}M{1.2cm}M{1cm}} \toprule

		\multicolumn{1}{c}{\textbf{Tool name}} & \textbf{Open Source} & \textbf{Long-Term Investment} & \textbf{Market} & \textbf{Stochastic Inputs} & \textbf{Country Generalisability} \\ \midrule
		SEPIA \cite{Harp2000}  & \checkmark           & $\times$                             & \checkmark      & Demand                     & \checkmark                        \\ 
		EMCAS ~\cite{Conzelmann}   & $\times$                    & \checkmark                    & \checkmark      & Outages                    & \checkmark                        \\ 
		NEMSIM ~\cite{Batten2006}  & ?              & \checkmark                    & \checkmark      & $\times$                          & $\times$                                 \\ 
		AMES  ~\cite{Sun2007} & \checkmark           & $\times$                             & Day-ahead       & $\times$                          & $\times$                                 \\ 
		GAPEX  ~\cite{Cincotti2013} & ?              & $\times$                             & Day-ahead       & $\times$                          & \checkmark                        \\ 
		PowerACE \cite{Rothengatter2007} & $\times$                    & \checkmark                    & \checkmark      & Outages Demand             & \checkmark                        \\ 

		EMLab ~\cite{Chappin2017}  & \checkmark           & \checkmark                    & Futures         & $\times$                          & \checkmark                        \\ 
		MACSEM  ~\cite{Praca2003}  & ?              & $\times$                             & \checkmark      & $\times$                          & \checkmark                        \\ 
		ElecSim                                  & \checkmark           & \checkmark                    & Futures         & \checkmark                 & \checkmark                        \\ \hline
	\end{tabular}
	\caption{Features of electricity market ABM tools.}
	\label{table:tool_comparison}
	\vskip -1cm
\end{table}

A number of ABM tools have emerged over the years to model electricity markets: SEPIA~\cite{Harp2000}, EMCAS~\cite{Conzelmann}, NEMSIM~\cite{Batten2006}, AMES~\cite{Sun2007}, GAPEX~\cite{Cincotti2013}, PowerACE~\cite{Rothengatter2007}, EMLab~\cite{Chappin2017} and MACSEM ~\cite{Praca2003}. Table \ref{table:tool_comparison} shows that these do not suit the needs of an open source, long-term market model. We will demonstrate that Monte-Carlo sampling of parameters is also required to increase realism.

There have been a number of recent studies using ABMs which focus on electricity markets, however they often utilize ad-hoc tools which are designed for a particular application \cite{Saxena2019, hadar2019, Kunzel2018}. ElecSim, however, has been built for re-use and reproducibility. The survey \cite{Weidlich2008} cites that many of these tools do not release source code or parameters, which is a problem that ElecSim seeks to address.

Table \ref{table:tool_comparison} contains six columns: tool name, whether the tool is open source or not, whether they model long-term investment in electricity infrastructure, and the markets they model. We determine how the stochasticity of real life is modelled, and determine whether the model is generalisable to different countries.

An open source toolkit is important for reproducibility, transparency and lowering barriers to entry. It enables users to expand the model to their requirements and respective country. The modelling of long-term investment enables scenarios to emerge, and enable users to model investment behaviour. We demonstrate that the use of a Monte-Carlo method improves results.

SEPIA \cite{Harp2000} is a discrete event ABM which utilises Q-learning to model the bids made by GenCos. SEPIA models plants as being always on, and does not have an independent system operator (ISO), which in an electricity market, is an independent non-profit organization for coordinating and controlling of regular operations of the electric power system and market \cite{Zhou2007}. SEPIA does not model a spot market, instead focusing on bilateral contracts. As opposed to this, ElecSim has been designed with a merit-order, spot market in mind. As shown in Table \ref{table:tool_comparison}, SEPIA does not include a long-term investment mechanism. 

EMCAS ~\cite{Conzelmann} is a closed source ABM. EMCAS investigates the interactions between physical infrastructures and economic behaviour of agents. However, ElecSim focuses on the dynamics of the market, and provides a simplified, transparent model of market operation, whilst maintaining robustness of results.

NEMSIM \cite{Grozev2005} is an ABM that represents Australia's National Electricity Market (NEM). Participants are able to grow and change over time using learning algorithms. NEMSIM is non-generalisable to other electricity markets, unlike ElecSim.

AMES ~\cite{Sun2007} is an ABM specific to the US Wholesale Power Market Platform and therefore not generalizable for other countries. GAPEX \cite{Cincotti2013} is an ABM framework for modelling and simulating power exchanges . GAPEX utilises an enhanced version of the reinforcement technique Roth-Erev \cite{RothAE1995} to consider the presence of affine total cost functions. However, neither of these model the long-term dynamics for which ElecSim is designed.

PowerACE ~\cite{Rothengatter2007} is a closed source ABM of electricity markets that integrates short-term daily electricity trading and long-term investment decisions. PowerACE models the spot market, forward market and a carbon market. Similarly to ElecSim, PowerACE initialises GenCos with each of their power plants. However, as can be seen in Table \ref{table:tool_comparison}, unlike ElecSim, PowerACE does not take into account stochasticity of price risks in electricity markets ~\cite{Most2010}.

EMLab ~\cite{Chappin2017} is an open-source ABM toolkit for the electricity market. Like PowerACE, EMLab models an endogenous carbon market, however, they both differ from ElecSim by not taking into account stochasticity in the electricity markets, such as in outages, fuel prices and operating costs. After correspondence with the authors, however, we were unable to run the current version.

MACSEM \cite{Praca2003} has been used to probe the effects of market rules and conditions by testing different bidding strategies. MACSEM does not model long term investments or stochastic inputs.

As can be seen from Table \ref{table:tool_comparison}, none of the tools fill each of the characteristics we have defined. We therefore propose ElecSim to contribute an open source, long-term, stochastic investment model.

\section{ElecSim Architecture} \label{Model}

\begin{figure}
	\centering
	\includegraphics[width=0.85\linewidth]{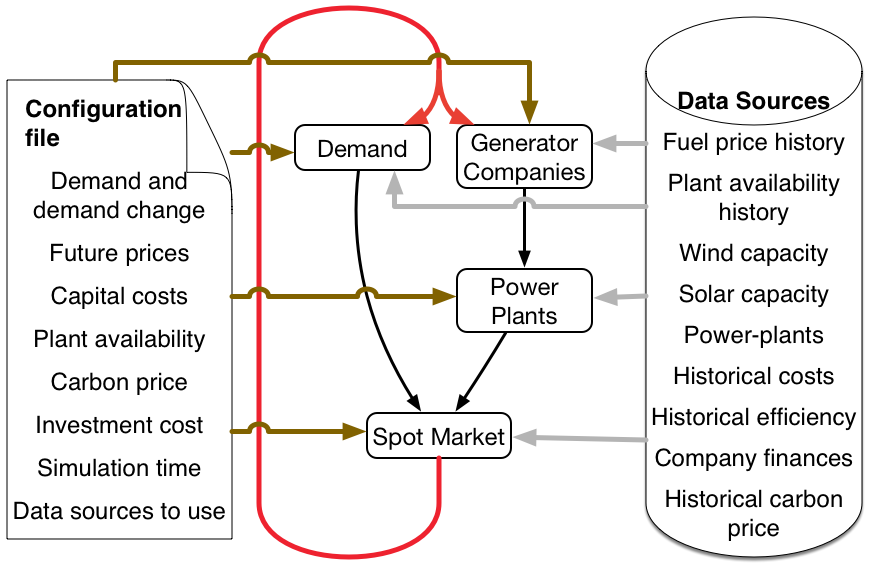}
	\caption{High level overview.}
	\label{fig:systemoverview}
	\vskip -6.9mm
\end{figure}

\begin{figure*}
	\centering
	\includegraphics[width=0.8\linewidth]{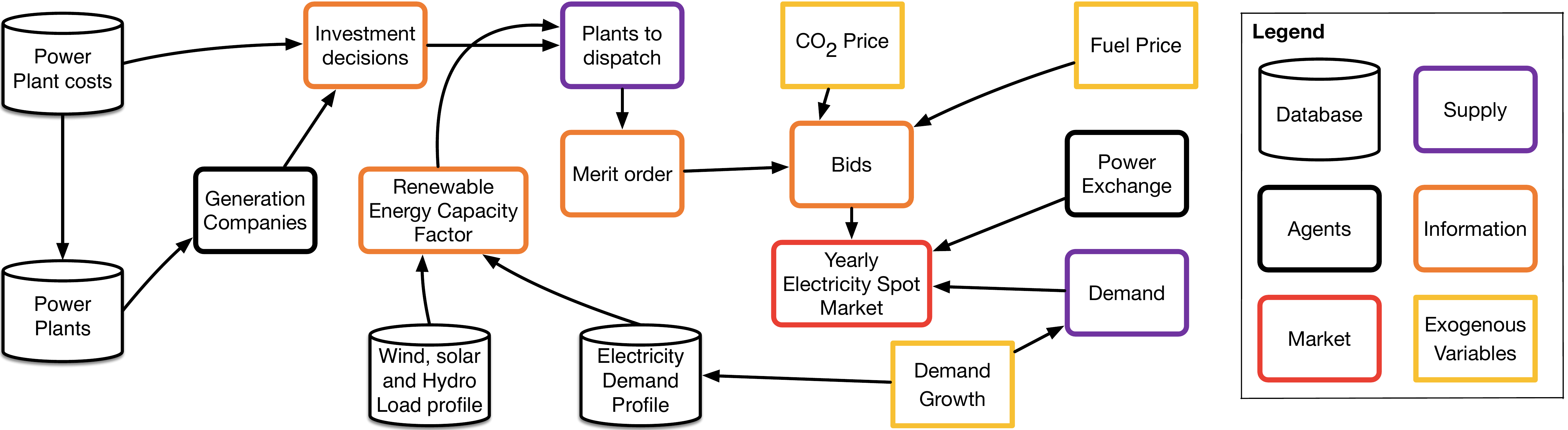}
	\caption{ElecSim simulation overview}
	\label{fig:lowlevelsystem}
\end{figure*}

ElecSim is made up of five fundamental parts: the agents, which are split up into demand and GenCos; power plants; a Power Exchange, which controls an electricity spot market; and the data for parametrisation. A schematic of ElecSim is displayed in Figure \ref{fig:systemoverview}.

\textit{Data parametrisation.} ElecSim contains a configuration file and a collection of data sources for parametrisation. These data sources contain information such as historical fuel prices, historical plant availability, wind and solar capacity.

The configuration file allows for rapid changes to test different hypothesis and scenarios, and points to the different data sources. The configuration file enables one to change the demand growth and shape, future fuel and carbon prices, capital costs, plant availability, investment costs and simulation time.

\textit{Demand Agent.} The demand agent is a simplified representation of aggregated demand in a country. The demand is represented as a load duration curve (LDC). \vphantom{\color{red}An example load duration curve for a year is demonstrated in Figure \ref{fig:loaddurationcurve}.} An LDC is an arrangement of all load levels in descending order of magnitude. \vphantom{\color{red}, where the lowest segment demand demonstrates baseload, and the highest segment represents peak demand.} Each year, the demand agent changes each of the LDC segments proportionally.


As per Chappin \textit{et al.} \cite{Chappin2017}, we modelled the LDC of electricity demand with twenty segments. Twenty segments enabled us to capture the variation in demand throughout the year to a high degree of accuracy, whilst reducing computational complexity.

\textit{Generation Company Agents.} The GenCos have two main functions. Investing in power plants and making bids to sell their generation capacity. We will first focus on the buying and selling of electricity, and then cover the investment algorithm.

The power exchange runs every year, accepting the lowest bids until supply meets demand. Once this condition is met, the spot price or system marginal price (SMP) is paid to all generators regardless of their initial bid. Generators are motivated to bid their SRMC, to ensure that their generator is being utilised, and reduce the risk of overbidding.

\textit{Investment.} Investment in power plants is made based upon a net present value (NPV) calculation. NPV is a summation of the present value of a series of present and future cash flow. NPV provides a method for evaluating and comparing investments with cash flows spread over many years, making it suited for evaluating power plants which have a long lifetime.  \vphantom{\color{red}NPV is based upon the fact that current cash flow is worth more than future cash flow. This is due to the fact that money today can be invested and have a rate of return. This means that, for example \$50,000 today is worth more than \$50,000 in 10 years time. The value in which future cash flow is worth less than present cash flow is discounted by the discount rate.}

Equation \ref{eq:npv_eq} is the calculation of NPV, where $t$ is the year of the cash flow, $i$ is the discount rate, $N$ is total number of periods, or lifetime of power plant, and $R_t$ is the net cash flow at time $t$.
\begin{equation} \label{eq:npv_eq}
NPV(i, N) = \sum_{t=0}^{N}\frac{R_t}{(1+t)^t}
\end{equation}
A discount rate set by a GenCo's weighted average cost of capital (WACC) is often used \cite{KincheloeStephenC1990TWAC}. WACC is the rate that a company is expected to pay on average for its stock and debt. Therefore to achieve a positive NPV, an income larger than the WACC is required. However, a higher WACC is often selected to adjust for varying risk profiles, opportunity costs and rates of return. To account for these differences we sample from a Gaussian distribution, giving us sufficient variance whilst deviating from the expected price.

To calculate the NPV, future market conditions must be considered. For this, each GenCo forecasts $N$ years into the future, which we assume is representative of the lifetime of the plant. As in the real world, GenCos have imperfect information, and therefore must forecast expected demand, fuel prices, carbon price and electricity sale price. This is achieved by fitting functions to historical data. Each GenCo is different in that they will use differing historical time periods of data for forecasting.

Fuel and carbon price are forecast using linear regression. Demand, however, is forecast using an exponential function, which considers compounded growth. Linear regression is used if an exponential function is found to be sub-optimal.

This forecasted data is then used to simulate a market $N$ years into the future using the electricity market algorithm. We simulate a market based on the expected bids -- based on SRMC -- that every operating power plant will make. This includes the removal of plants that will be past their operating period, and the introduction of plants that are in construction or pre-development stages. 

There may be scenarios where demand is forecast to grow significantly, and limited investments have yet been made to meet that demand. The expected price, would be that of lost load. Lost load is defined as the price customers would be willing to pay to avoid disruption in their electricity supply. To avoid GenCos from estimating large profits, and under the assumption that further power plant investments will be made, the lost load price is replaced with a predicted electricity price using linear regression based on prices at lower points of the demand curve. If zero segments of demand are met, then the  lost load price is used to encourage investment. 

Once this data has been forecasted\vphantom{Once expected fuel prices, carbon price, discount rate, and expected sale price of electricity are all forecast}, the NPV can be calculated. GenCos must typically provide a certain percentage of upfront capital, with the rest coming from investors in the form of stock and shares or debt (WACC). The percentage of upfront capital can be customised by the user in the configuration file. The GenCos then invest in the power plants with the highest NPV.

\textit{Power Plant Parameters.}\label{ssssec:powerplantparameters} Costs form an important element of markets and investment, and publicly available data for power plant costs for individual countries can be scarce. Thus, extrapolation and interpolation is required to estimate costs for power plants of differing sizes, types and years of construction.

Users are able to initialise costs relevant to their particular country by providing detailed cost parameters. They can also provide an average cost per MWh produced over the lifetime of a plant, known as levelised cost of electricity (LCOE).

The parameters used to initialise the power plants are detailed in this section. Periods have units of years and costs in \textsterling/MW unless otherwise stated: Efficiency ($\eta$) is defined as the percentage of energy from fuel that is converted into electrical energy (\%). Operating period ($OP$) is the total period in which a power plant is in operation. Pre-development period ($P_D$) and pre-development costs ($P_C$) include the time and costs for pre-licensing, technical and design, as well as costs incurred due to regulatory, licensing and public enquiry. The construction period ($C_D$) and construction costs ($C_C$) are incurred during the development of the plant, excluding network connections. The infrastructure costs ($I_C$) are the costs incurred by the developer in connecting the plant to the electricity or gas grid (\textsterling). Fixed operation \& maintenance costs ($F_C$) are costs incurred in operating the plant that do not vary based on output. Variable operation \& maintenance ($V_C$) costs are incurred in operating the plant that depend on generator output \cite{Ltd2016}.


Precise data is not available for every plant size. Linear interpolation is used to estimate individual prices between known points. When the plant to be estimated falls outside of the range of known data points, the closest power plant is used. We experimented with extrapolation but this would often lead to unrealistic costs. 

If specific parameters are not known the LCOE can be used for parameter estimation, through the use of linear optimisation. Constraints can be set by the user, enabling, for example, varying operation and maintenance costs per country as a fraction of LCOE.

To fully parametrise power plants, availability and capacity factors are required. Availability is the percentage of time that a power plant can produce electricity. This can be reduced by forced or planned outages. We integrate historical data to model improvements in reliability over time.

The capacity factor is the actual electrical energy produced over a given time period divided by the maximum possible electrical energy it could have produced. The capacity factor can be impacted by regulatory constraints, market forces and resource availability. For example, higher capacity factors are common for photovoltaics in the summer, and lower in winter. 

To model the intermittency of wind and solar power we allow them to contribute only a certain percentage of their total capacity (nameplate capacity) for each load segment. This percentage is based upon empirical wind and solar capacity factors. In this calculation we consider the correlation between demand and renewable resources. We are unable to model short-term storage due to ElecSim taking a single time-step per year. 

When initialised, $V_C$ is selected from a uniform distribution, with the ability for the user to set maximum percentage increase or decrease. A uniform distribution was chosen to capture the large deviations that can occur in $V_C$, especially over a long time period. \vphantom{By doing this, the variance in costs between individual power plants for processes such as preventative and corrective maintenance, labour costs and skill, health and safety and chance are different per plant instant.}

Fuel price is controlled by the user, however, there is inherent volatility in fuel price. To take into account this variability, an ARIMA \cite{ARIMA} model was fit to historical gas and coal price data. The standard deviation of the residuals was used to model the variance in price that a GenCo will buy fuel in a given year. This considers differences in chance and hedging strategies.


Figure \ref{fig:lowlevelsystem} demonstrates the simulation and how it co-ordinates runs. The world contains data and brings together GenCos, the Power Exchange and demand. The investment decisions are based on future demand and costs, which in turn influence bids made.

Exogenous variables include fuel and \ce{CO2} prices as well as demand growth. Once the data is initialised, the world calls on the Power Exchange to operate the yearly electricity spot market. The world also settles the accounts of the GenCos, by paying bids, and removing operating and capital costs as well as loans and dividends.

\section{Validation and Performance}\label{Validation and Performance}
\paragraph{Validation}

 Validation of models is important to ascertain that the output is accurate. However, it should be noted that these long-term simulations are not predictions of the future, rather possible outcomes based upon certain assumptions. Jager posits that a certain outcome or development path, captured by empirical data, might have developed in a completely different direction due to chance. However, the processes that emerge from a model should be realistic and in keeping with expected behaviour \cite{Jager2006a}.

\begin{figure}
	\begin{center}
		\includegraphics[width=0.35\textwidth]{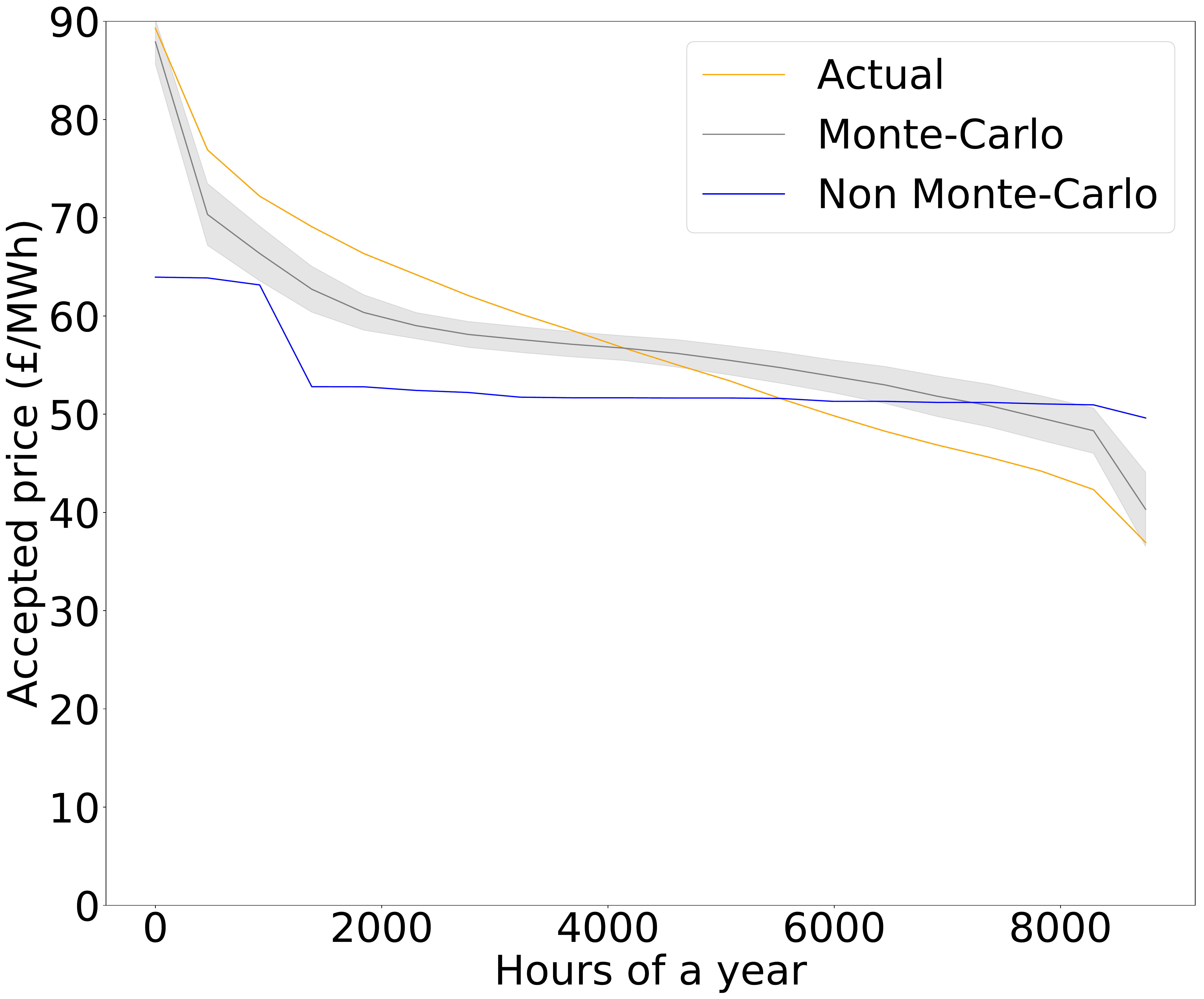}
		\caption{Price duration curve which compares real electricity prices to those paid in ElecSim (2018).}
		\label{fig:price_duration_curve}
	\end{center}
\end{figure}

We begin by comparing the price duration curve in the year 2018. Figure \ref{fig:price_duration_curve} shows the N2EX Day Ahead Auction Prices of the UK \cite{nordpool_2019}, the Monte-Carlo simulated electricity prices, and the non Monte-Carlo electricity price throughout the year 2018. Fuel prices varying throughout a year, as does $V_C$ and WACC. WACC is sampled from a Gaussian distribution with a standard deviation of $\pm3$\%. $V_C$ is sampled from a uniform distribution between 30\% and 200\% of the mean $V_C$ price, whilst fuel price is sampled from the residuals of an ARIMA model fit on historical data. The N2EX Day Ahead Market is a day ahead market run by Nord Pool AS. Nord Pool AS runs the largest market for electrical energy in Europe, measured in volume traded and in market share \cite{nordpool_2019}.


\begin{table}[]
\small
\begin{tabular}{p{3cm}p{1.4cm}p{1cm}p{1.5cm}}
\hline
Metric & N2EX Day Ahead & ElecSim & Non Monte-Carlo \\ \hline
Avg. Price (\textsterling/MWh) & 57.49 & 57.52 & 53.39 \\
Std. dev (\textsterling/MWh) & - & 9.64 & - \\
MAE (\textsterling/MWh) & - & 3.97 & 8.35 \\
RMSE (\textsterling/MWh) & - & 4.41 & 10.2 \\ \hline
\end{tabular}
\caption{Validation performance metrics.}
\label{table:validation_metrics}
\vskip -10mm
\end{table}

We ran the initialisation of the model 40 times to capture the price variance. Outliers were removed as on a small number of occasions large jumps in prices at peak demand occurred which deviated from the mean. We did this, as although this does occur in real life, it occurs at a smaller fraction of the time than 5\% of the year (modelled LDC), therefore the results would be unreasonably skewed for the highest demand segment. 

Figure \ref{fig:price_duration_curve} demonstrates very little variance in the non-stochastic case. This is due to the fact that combined cycle gas turbines (CCGTs) set the spot price. These CCGTs have little variance between one another as they were calibrated using the same dataset. By adding stochasticity of fuel prices and operation and maintenance prices, a curve that more closely resembles the actual data occurs. The stochastic curve, however, does not perfectly fit the real data, which may be due to higher variance in fuel prices and historical differences in operation and maintenance costs between power plants. One method of improving this would be fitting the data used to parametrise to the curve.

Table \ref{table:validation_metrics} shows performance metrics of the stochastic and non-stochastic runs versus the actual price duration curve . The stochastic implementation, improves the mean absolute error (MAE) of the non-stochastic case by $52.5\%$.

By observing the processes that emerge from the long-term scenarios, we can see that carbon price and investment in renewable generation are positively correlated, as would be expected. The highest NPV calculations were for onshore wind and CCGT plants. This is realistic for the United Kingdom, where subsidies are required for other forms of generation such as coal and nuclear.

\paragraph{Performance}

 We used 	

\begin{figure}
	\centering
	\includegraphics[width=0.7\linewidth]{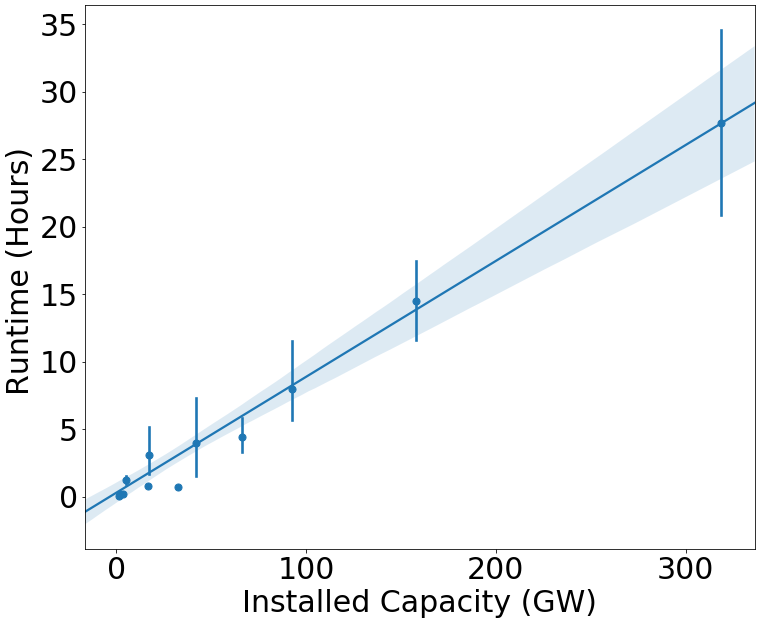}
	\caption{Run times of different sized countries.}
	\label{fig:timingplot}
	\vskip -0.5cm
\end{figure}

Figure \ref{fig:timingplot} shows the running time for ElecSim with varying installed capacity. We varied demand between 2GW and 320GW to see the effect of different sized countries on running time. The makeup of the electricity mix was achieve through stratified sampling of the UK electricity mix. The results show a linear time complexity.


\section{Scenario Testing}\label{Scenario Testing}

Here we present example scenario runs using ElecSim. We vary the carbon tax and grow or reduce total electricity demand. This enables us to observe the effects of carbon tax on investment. In this paper we have presented scenarios where electricity demand decreases 1\% per year, due to the recent trend in the UK.

\begin{figure}
	\centering
	\begin{subfigure}{.8\linewidth}
		\includegraphics[width=\linewidth]{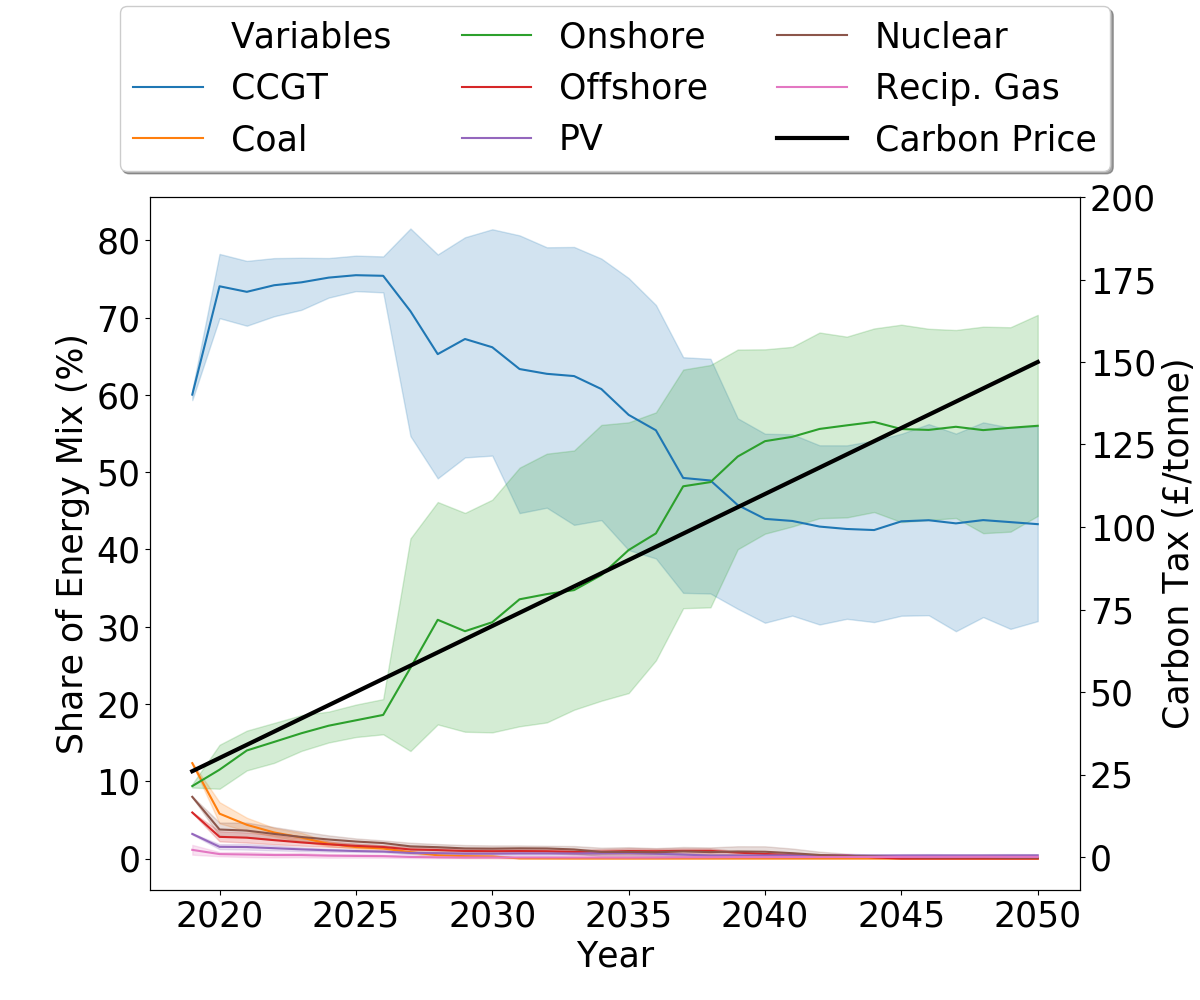}
		\caption{\textsterling26 to \textsterling150 linearly increasing carbon tax.}
		\label{fig:demand99carbon18}
	\end{subfigure}
	\begin{subfigure}{.8\linewidth}
		\includegraphics[width=\linewidth]{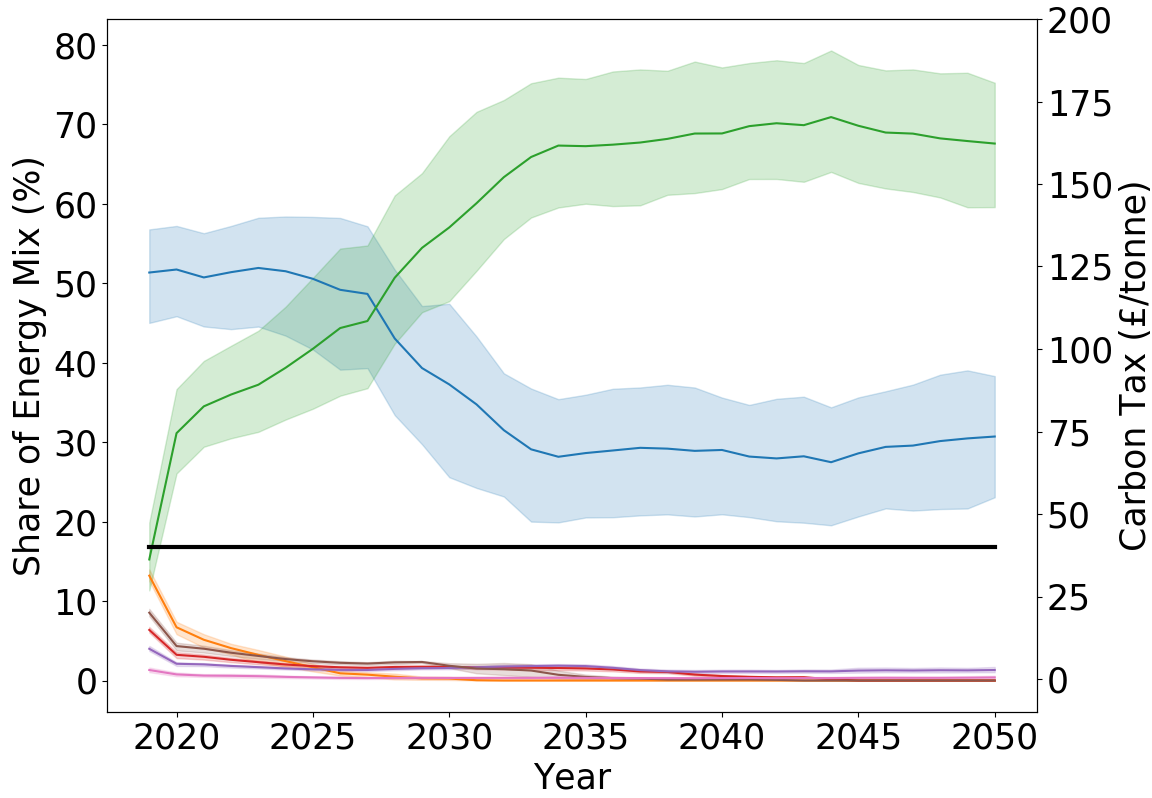}
		\caption{{\textsterling40 carbon tax}}
		\label{fig:demand99carbon40}
	\end{subfigure}
	\caption{Scenarios with varying carbon taxes and decreasing demand (-1\%/year)}
\end{figure}

For the first scenario run displayed, we have approximated the predictions by the UK Government, where carbon tax increases linearly from \textsterling18 to \textsterling200 by 2050 \cite{Department2016}. Figure \ref{fig:demand99carbon18} demonstrates a significant increase in gas turbines in the first few years, followed by a decrease, with onshore wind increasing.

Figure \ref{fig:demand99carbon40} displays a run with a \textsterling40 carbon tax. This run demonstrates a higher share of onshore wind than in the previous scenario. 

These runs demonstrate that a consistent, but relatively low carbon tax can have a larger impact in the uptake of renewable energy than increasing carbon tax over a long time frame. We hypothesise that an early carbon tax affects the long-term dynamics of the market for many years. We, therefore, suggest early action on carbon tax to transition to a low-carbon energy supply

\section{Conclusions}\label{Conclusion}

Liberalised electricity markets with many heterogenous players are suited to be modelled with ABMs. ABMs incorporate imperfect information as well as heterogeneous actors. ElecSim models imperfect information through forecasting of electricity demand and future fuel and electricity prices. This leads to agents taking risk on their investments, and model market conditions more realistically.

We demonstrated that increasing carbon tax can lead to an increase in investment of low-carbon technologies. We showed that early decisions have a long-term impact on the energy mix. 

Our future work includes comparing agent-learning techniques, using multi-agent reinforcement learning algorithms to allow agents to learn in a non-static environment. We propose the integration of a higher temporal and spatial resolution to model changes in daily demand, as well as capacity factors by region, and transmission effects. This will allow us to model that demand is met at all times and not just on average. 

\FloatBarrier

%


%
\bibliographystyle{ACM-Reference-Format}
\bibliography{library,custombibtex}


\begin{thebibliography}{38}


\ifx \showCODEN    \undefined \def \showCODEN     #1{\unskip}     \fi
\ifx \showDOI      \undefined \def \showDOI       #1{#1}\fi
\ifx \showISBNx    \undefined \def \showISBNx     #1{\unskip}     \fi
\ifx \showISBNxiii \undefined \def \showISBNxiii  #1{\unskip}     \fi
\ifx \showISSN     \undefined \def \showISSN      #1{\unskip}     \fi
\ifx \showLCCN     \undefined \def \showLCCN      #1{\unskip}     \fi
\ifx \shownote     \undefined \def \shownote      #1{#1}          \fi
\ifx \showarticletitle \undefined \def \showarticletitle #1{#1}   \fi
\ifx \showURL      \undefined \def \showURL       {\relax}        \fi
\providecommand\bibfield[2]{#2}
\providecommand\bibinfo[2]{#2}
\providecommand\natexlab[1]{#1}
\providecommand\showeprint[2][]{arXiv:#2}

\bibitem[\protect\citeauthoryear{Batten and Grozev}{Batten and Grozev}{2006}]%
        {Batten2006}
\bibfield{author}{\bibinfo{person}{David Batten} {and} \bibinfo{person}{George
  Grozev}.} \bibinfo{year}{2006}\natexlab{}.
\newblock \showarticletitle{{NEMSIM: Finding Ways to Reduce Greenhouse Gas
  Emissions Using Multi-Agent Electricity Modelling}}.
\newblock \bibinfo{journal}{\emph{Complex Science for a Complex World}}
  (\bibinfo{year}{2006}), \bibinfo{pages}{227--252}.
\newblock


\bibitem[\protect\citeauthoryear{B{\"{o}}hringer}{B{\"{o}}hringer}{1998}]%
        {Bohringer1998}
\bibfield{author}{\bibinfo{person}{Christoph B{\"{o}}hringer}.}
  \bibinfo{year}{1998}\natexlab{}.
\newblock \showarticletitle{{The synthesis of bottom-up and top-down in energy
  policy modeling}}.
\newblock \bibinfo{journal}{\emph{Energy Economics}} \bibinfo{volume}{20},
  \bibinfo{number}{3} (\bibinfo{year}{1998}), \bibinfo{pages}{233--248}.
\newblock
\showISBNx{0140-9883}
\showISSN{01409883}


\bibitem[\protect\citeauthoryear{BP}{BP}{2018}]%
        {BP2018}
\bibfield{author}{\bibinfo{person}{BP}.} \bibinfo{year}{2018}\natexlab{}.
\newblock \showarticletitle{{BP Statistical Review of World Energy}}.
\newblock  (\bibinfo{year}{2018}), \bibinfo{pages}{1--56}.
\newblock
\showISSN{0300-3604}


\bibitem[\protect\citeauthoryear{Chappin, de~Vries, Richstein, Bhagwat,
  Iychettira, and Khan}{Chappin et~al\mbox{.}}{2017}]%
        {Chappin2017}
\bibfield{author}{\bibinfo{person}{Emile~J.L. Chappin},
  \bibinfo{person}{Laurens~J. de Vries}, \bibinfo{person}{Joern~C. Richstein},
  \bibinfo{person}{Pradyumna Bhagwat}, \bibinfo{person}{Kaveri Iychettira},
  {and} \bibinfo{person}{Salman Khan}.} \bibinfo{year}{2017}\natexlab{}.
\newblock \showarticletitle{{Simulating climate and energy policy with
  agent-based modelling: The Energy Modelling Laboratory (EMLab)}}.
\newblock \bibinfo{journal}{\emph{Environmental Modelling and Software}}
  \bibinfo{volume}{96} (\bibinfo{year}{2017}), \bibinfo{pages}{421--431}.
\newblock
\showISSN{13648152}


\bibitem[\protect\citeauthoryear{Cincotti, Gallo, and Berkeley}{Cincotti
  et~al\mbox{.}}{2013}]%
        {Cincotti2013}
\bibfield{author}{\bibinfo{person}{Silvano Cincotti}, \bibinfo{person}{Giulia
  Gallo}, {and} \bibinfo{person}{Lawrence Berkeley}.}
  \bibinfo{year}{2013}\natexlab{}.
\newblock \showarticletitle{{The Genoa Artificial Power-Exchange The Genoa
  artificial power-exchange}}.
\newblock  \bibinfo{number}{January} (\bibinfo{year}{2013}).
\newblock
\showISBNx{9783642369070}


\bibitem[\protect\citeauthoryear{Conzelmann, Boyd, Koritarov, and
  Veselka}{Conzelmann et~al\mbox{.}}{[n. d.]}]%
        {Conzelmann}
\bibfield{author}{\bibinfo{person}{G. Conzelmann}, \bibinfo{person}{G. Boyd},
  \bibinfo{person}{V. Koritarov}, {and} \bibinfo{person}{T. Veselka}.}
  \bibinfo{year}{[n. d.]}\natexlab{}.
\newblock \showarticletitle{{Multi-agent power market simulation using EMCAS}}.
\newblock \bibinfo{journal}{\emph{IEEE Power Engineering Society General
  Meeting, 2005}} (\bibinfo{year}{[n. d.]}), \bibinfo{pages}{917--922}.
\newblock
\showISBNx{0-7803-9157-8}
\showISSN{1932-5517; 0-7803-9156-X}
\urldef\tempurl%
\url{http://ieeexplore.ieee.org/document/1489271/}
\showURL{%
\tempurl}


\bibitem[\protect\citeauthoryear{Cook, Nuccitelli, Green, Richardson, Winkler,
  Painting, Way, Jacobs, and Skuce}{Cook et~al\mbox{.}}{2013}]%
        {Cook2013}
\bibfield{author}{\bibinfo{person}{John Cook}, \bibinfo{person}{Dana
  Nuccitelli}, \bibinfo{person}{Sarah~A Green}, \bibinfo{person}{Mark
  Richardson}, \bibinfo{person}{B{\"{a}}rbel Winkler}, \bibinfo{person}{Rob
  Painting}, \bibinfo{person}{Robert Way}, \bibinfo{person}{Peter Jacobs},
  {and} \bibinfo{person}{Andrew Skuce}.} \bibinfo{year}{2013}\natexlab{}.
\newblock \showarticletitle{{Quantifying the consensus on anthropogenic global
  warming in the scientific literature}}.
\newblock \bibinfo{journal}{\emph{Environ. Res. Lett}}  \bibinfo{volume}{8}
  (\bibinfo{year}{2013}), \bibinfo{pages}{24024--7}.
\newblock
\showISSN{1748-9326}


\bibitem[\protect\citeauthoryear{{Department for Business Energy {\&}
  Industrial Strategy}}{{Department for Business Energy {\&} Industrial
  Strategy}}{2016}]%
        {Department2016}
\bibfield{author}{\bibinfo{person}{{Department for Business Energy {\&}
  Industrial Strategy}}.} \bibinfo{year}{2016}\natexlab{}.
\newblock \showarticletitle{{Electricity Generation Costs}}.
\newblock  \bibinfo{number}{November} (\bibinfo{year}{2016}).
\newblock


\bibitem[\protect\citeauthoryear{Fishbone and Abilock}{Fishbone and
  Abilock}{1981}]%
        {Fishbone1981}
\bibfield{author}{\bibinfo{person}{Leslie~G. Fishbone} {and}
  \bibinfo{person}{Harold Abilock}.} \bibinfo{year}{1981}\natexlab{}.
\newblock \showarticletitle{{Markal, a linear programming model for energy
  systems analysis: Technical description of the bnl version}}.
\newblock \bibinfo{journal}{\emph{International Journal of Energy Research}}
  \bibinfo{volume}{5}, \bibinfo{number}{4} (\bibinfo{year}{1981}),
  \bibinfo{pages}{353--375}.
\newblock
\showISBNx{1099-114X}
\showISSN{1099114X}


\bibitem[\protect\citeauthoryear{Gargiulo and Brian}{Gargiulo and
  Brian}{2013}]%
        {Gargiulo2013}
\bibfield{author}{\bibinfo{person}{Maurizio Gargiulo} {and} \bibinfo{person}{O
  Brian}.} \bibinfo{year}{2013}\natexlab{}.
\newblock \showarticletitle{{Long-term energy models : Principles ,
  characteristics , focus ,}}.
\newblock \bibinfo{journal}{\emph{WIREs Energy and Environment}}
  \bibinfo{volume}{2}, \bibinfo{number}{April} (\bibinfo{year}{2013}).
\newblock


\bibitem[\protect\citeauthoryear{Group}{Group}{2019}]%
        {nordpool_2019}
\bibfield{author}{\bibinfo{person}{Nord~Pool Group}.}
  \bibinfo{year}{2019}\natexlab{}.
\newblock \showarticletitle{N2EX Day Ahead Auction Prices}.
\newblock \bibinfo{journal}{\emph{Nordpoolgroup.com}} (\bibinfo{year}{2019}).
\newblock


\bibitem[\protect\citeauthoryear{Grozev, Batten, Anderson, Lewis, Mo, and
  Katzfey}{Grozev et~al\mbox{.}}{2005}]%
        {Grozev2005}
\bibfield{author}{\bibinfo{person}{George Grozev}, \bibinfo{person}{David
  Batten}, \bibinfo{person}{Miles Anderson}, \bibinfo{person}{Geoff Lewis},
  \bibinfo{person}{John Mo}, {and} \bibinfo{person}{Jack Katzfey}.}
  \bibinfo{year}{2005}\natexlab{}.
\newblock \showarticletitle{{NEMSIM: agent-based simulator for Australia's
  national electricity market}}.
\newblock \bibinfo{journal}{\emph{SimTecT 2005 Conference Proceedings, Sydney,
  Australia}} (\bibinfo{year}{2005}).
\newblock


\bibitem[\protect\citeauthoryear{Hadar and Hartmann}{Hadar and
  Hartmann}{2019}]%
        {hadar2019}
\bibfield{author}{\bibinfo{person}{Adam Hadar} {and} \bibinfo{person}{Balint
  Hartmann}.} \bibinfo{year}{2019}\natexlab{}.
\newblock \showarticletitle{{Simulation of cross-border capacity allocation}}.
\newblock \bibinfo{journal}{\emph{1st IEEE Student Conference on Electric
  Machines and Systems, SCEMS 2018}} (\bibinfo{year}{2019}),
  \bibinfo{pages}{1--6}.
\newblock
\showISBNx{9781538673485}


\bibitem[\protect\citeauthoryear{Hall and Buckley}{Hall and Buckley}{2016}]%
        {Hall2016}
\bibfield{author}{\bibinfo{person}{Lisa M~H Hall} {and}
  \bibinfo{person}{Alastair~R Buckley}.} \bibinfo{year}{2016}\natexlab{}.
\newblock \showarticletitle{{A review of energy systems models in the UK :
  Prevalent usage and categorisation}}.
\newblock \bibinfo{journal}{\emph{Applied Energy}}  \bibinfo{volume}{169}
  (\bibinfo{year}{2016}), \bibinfo{pages}{607--628}.
\newblock
\showISSN{0306-2619}


\bibitem[\protect\citeauthoryear{Harp, Wollenberg, and Samad}{Harp
  et~al\mbox{.}}{2000}]%
        {Harp2000}
\bibfield{author}{\bibinfo{person}{By~Steven~A Harp}, \bibinfo{person}{Bruce~F
  Wollenberg}, {and} \bibinfo{person}{Tariq Samad}.}
  \bibinfo{year}{2000}\natexlab{}.
\newblock \showarticletitle{{SEPIA: A Simulator for Electric Power Industry
  Agents}}.
\newblock  \bibinfo{number}{August} (\bibinfo{year}{2000}),
  \bibinfo{pages}{53--69}.
\newblock


\bibitem[\protect\citeauthoryear{Heaps}{Heaps}{2016}]%
        {Heaps2016}
\bibfield{author}{\bibinfo{person}{C.G. Heaps}.}
  \bibinfo{year}{2016}\natexlab{}.
\newblock \showarticletitle{Long-range Energy Alternatives Planning (LEAP)
  system}.
\newblock  (\bibinfo{year}{2016}).
\newblock


\bibitem[\protect\citeauthoryear{IEA}{IEA}{2015}]%
        {IEA2015}
\bibfield{author}{\bibinfo{person}{IEA}.} \bibinfo{year}{2015}\natexlab{}.
\newblock \showarticletitle{{Projected Costs of Generating Electricity}}.
\newblock  (\bibinfo{year}{2015}), \bibinfo{pages}{215}.
\newblock
\showISBNx{9789264244160}


\bibitem[\protect\citeauthoryear{IRENA}{IRENA}{2018}]%
        {IRENA2018}
\bibfield{author}{\bibinfo{person}{IRENA}.} \bibinfo{year}{2018}\natexlab{}.
\newblock \bibinfo{booktitle}{\emph{{Renewable Power Generation Costs in 2017.
  IRENA - International Renewable Energy Agency}}}.
\newblock 160 pages.
\newblock
\showISBNx{978-92-9260-040-2}
\showISSN{1476-4687}


\bibitem[\protect\citeauthoryear{Jager}{Jager}{2006}]%
        {Jager2006a}
\bibfield{author}{\bibinfo{person}{Wander Jager}.}
  \bibinfo{year}{2006}\natexlab{}.
\newblock \showarticletitle{{Simulating consumer behaviour: a perspective}}.
\newblock \bibinfo{journal}{\emph{Environmental Policy and Modelling in
  Evolutionary Economics}} (\bibinfo{year}{2006}), \bibinfo{pages}{1--28}.
\newblock


\bibitem[\protect\citeauthoryear{Keles, Jochem, Mckenna, Ruppert, and
  Fichtner}{Keles et~al\mbox{.}}{2017}]%
        {Keles2017}
\bibfield{author}{\bibinfo{person}{Dogan Keles}, \bibinfo{person}{Patrick
  Jochem}, \bibinfo{person}{Russell Mckenna}, \bibinfo{person}{Manuel Ruppert},
  {and} \bibinfo{person}{Wolf Fichtner}.} \bibinfo{year}{2017}\natexlab{}.
\newblock \showarticletitle{{Meeting the Modeling Needs of Future Energy
  Systems}}.
\newblock \bibinfo{journal}{\emph{Energy Technology}} (\bibinfo{year}{2017}),
  \bibinfo{pages}{1007--1025}.
\newblock


\bibitem[\protect\citeauthoryear{Kincheloe}{Kincheloe}{1990}]%
        {KincheloeStephenC1990TWAC}
\bibfield{author}{\bibinfo{person}{Stephen~C Kincheloe}.}
  \bibinfo{year}{1990}\natexlab{}.
\newblock \showarticletitle{The Weighted Average Cost Of Capital - The Correct
  Discount}.
\newblock \bibinfo{journal}{\emph{The Appraisal journal.}}
  \bibinfo{volume}{58}, \bibinfo{number}{1} (\bibinfo{year}{1990}).
\newblock
\showISSN{0003-7087}


\bibitem[\protect\citeauthoryear{K{\"{u}}nzel, Gmbh, Kg, Klumpp, Gmbh, and
  Kg}{K{\"{u}}nzel et~al\mbox{.}}{2018}]%
        {Kunzel2018}
\bibfield{author}{\bibinfo{person}{Thomas K{\"{u}}nzel},
  \bibinfo{person}{Fichtner Gmbh}, \bibinfo{person}{Co Kg},
  \bibinfo{person}{Florian Klumpp}, \bibinfo{person}{Fichtner Gmbh}, {and}
  \bibinfo{person}{Co Kg}.} \bibinfo{year}{2018}\natexlab{}.
\newblock \showarticletitle{{Bidding Strategies for Flexible and Inflexible
  Generation in a Power Market Simulation Model}}.
\newblock  (\bibinfo{year}{2018}), \bibinfo{pages}{532--537}.
\newblock
\showISBNx{9781450357678}


\bibitem[\protect\citeauthoryear{Law and Kelton}{Law and Kelton}{2000}]%
        {Law:603360}
\bibfield{author}{\bibinfo{person}{Averill~M Law} {and}
  \bibinfo{person}{David~W Kelton}.} \bibinfo{year}{2000}\natexlab{}.
\newblock \bibinfo{booktitle}{\emph{Simulation modeling and analysis; 3rd ed.}}
\newblock \bibinfo{publisher}{McGraw-Hill}, \bibinfo{address}{New York, NY}.
\newblock
\urldef\tempurl%
\url{http://cds.cern.ch/record/603360}
\showURL{%
\tempurl}


\bibitem[\protect\citeauthoryear{Ltd}{Ltd}{2016}]%
        {Ltd2016}
\bibfield{author}{\bibinfo{person}{LeighFisher Ltd}.}
  \bibinfo{year}{2016}\natexlab{}.
\newblock \bibinfo{title}{{Final Report: Electricity Generation Costs and
  Hurdle Rates}}.
\newblock
\newblock


\bibitem[\protect\citeauthoryear{Masson-Delmotte, Zhai, P{\"{o}}rtner, Roberts,
  Skea, Shukla, Pirani, Moufouma-Okia, P{\'{e}}an, Pidcock, Connors, Matthews,
  Chen, Zhou, Gomis, Lonnoy, Maycock, Tignor, and Waterfield}{Masson-Delmotte
  et~al\mbox{.}}{2018}]%
        {Masson-Delmotte2018}
\bibfield{author}{\bibinfo{person}{V Masson-Delmotte}, \bibinfo{person}{P
  Zhai}, \bibinfo{person}{H.O P{\"{o}}rtner}, \bibinfo{person}{D Roberts},
  \bibinfo{person}{J Skea}, \bibinfo{person}{P~R Shukla}, \bibinfo{person}{A
  Pirani}, \bibinfo{person}{W Moufouma-Okia}, \bibinfo{person}{C P{\'{e}}an},
  \bibinfo{person}{R Pidcock}, \bibinfo{person}{S Connors},
  \bibinfo{person}{J~B Matthews}, \bibinfo{person}{Y Chen}, \bibinfo{person}{X
  Zhou}, \bibinfo{person}{M~I Gomis}, \bibinfo{person}{E Lonnoy},
  \bibinfo{person}{T Maycock}, \bibinfo{person}{M Tignor}, {and}
  \bibinfo{person}{T Waterfield}.} \bibinfo{year}{2018}\natexlab{}.
\newblock \bibinfo{booktitle}{\emph{{IPCC Special Report 1.5 - Summary for
  Policymakers}}}.
\newblock
\showISBNx{9789291691432}
\showISSN{1476-4687}


\bibitem[\protect\citeauthoryear{M{\"{o}}st and Keles}{M{\"{o}}st and
  Keles}{2010}]%
        {Most2010}
\bibfield{author}{\bibinfo{person}{Dominik M{\"{o}}st} {and}
  \bibinfo{person}{Dogan Keles}.} \bibinfo{year}{2010}\natexlab{}.
\newblock \showarticletitle{{A survey of stochastic modelling approaches for
  liberalised electricity markets}}.
\newblock \bibinfo{journal}{\emph{European Journal of Operational Research}}
  \bibinfo{volume}{207}, \bibinfo{number}{2} (\bibinfo{year}{2010}),
  \bibinfo{pages}{543--556}.
\newblock
\showISBNx{0377-2217}
\showISSN{03772217}


\bibitem[\protect\citeauthoryear{OWPB}{OWPB}{2016}]%
        {OWPB2016}
\bibfield{author}{\bibinfo{person}{OWPB}.} \bibinfo{year}{2016}\natexlab{}.
\newblock \showarticletitle{{Transmission Costs for Offshore Wind Final Report
  April 2016}}.
\newblock  \bibinfo{number}{April} (\bibinfo{year}{2016}), \bibinfo{pages}{25}.
\newblock


\bibitem[\protect\citeauthoryear{Perloff}{Perloff}{2012}]%
        {Perloff2012}
\bibfield{author}{\bibinfo{person}{Jeffery~M. Perloff}.}
  \bibinfo{year}{2012}\natexlab{}.
\newblock \bibinfo{booktitle}{\emph{{Microeconomics Sixth Edition}}}.
\newblock 804 pages.
\newblock
\showISBNx{9780131392632}


\bibitem[\protect\citeauthoryear{Pra{\c{c}}a, Ramos, Vale, and
  Ascem}{Pra{\c{c}}a et~al\mbox{.}}{2003}]%
        {Praca2003}
\bibfield{author}{\bibinfo{person}{Isabel Pra{\c{c}}a}, \bibinfo{person}{Carlos
  Ramos}, \bibinfo{person}{Zita Vale}, {and} \bibinfo{person}{M Ascem}.}
  \bibinfo{year}{2003}\natexlab{}.
\newblock \showarticletitle{{MASCEM : A Multiagent Markets}}.
\newblock  (\bibinfo{year}{2003}).
\newblock


\bibitem[\protect\citeauthoryear{Ringler, Keles, and Fichtner}{Ringler
  et~al\mbox{.}}{2016}]%
        {Ringler2016a}
\bibfield{author}{\bibinfo{person}{Philipp Ringler}, \bibinfo{person}{Dogan
  Keles}, {and} \bibinfo{person}{Wolf Fichtner}.}
  \bibinfo{year}{2016}\natexlab{}.
\newblock \showarticletitle{{Agent-based modelling and simulation of smart
  electricity grids and markets - A literature review}}.
\newblock \bibinfo{journal}{\emph{Renewable and Sustainable Energy Reviews}}
  \bibinfo{volume}{57}, \bibinfo{number}{September} (\bibinfo{year}{2016}),
  \bibinfo{pages}{205--215}.
\newblock
\showISSN{18790690}


\bibitem[\protect\citeauthoryear{{Roth AE} and {Erev I}}{{Roth AE} and {Erev
  I}}{1995}]%
        {RothAE1995}
\bibfield{author}{\bibinfo{person}{{Roth AE}} {and} \bibinfo{person}{{Erev
  I}}.} \bibinfo{year}{1995}\natexlab{}.
\newblock \showarticletitle{{Learning in extensive-form games: Experimental
  data and simple dynamic models in the intermediate term}}.
\newblock \bibinfo{journal}{\emph{Games and economic behavior}}
  \bibinfo{volume}{8}, \bibinfo{number}{1} (\bibinfo{year}{1995}),
  \bibinfo{pages}{164--212}.
\newblock


\bibitem[\protect\citeauthoryear{Rothengatter}{Rothengatter}{2007}]%
        {Rothengatter2007}
\bibfield{author}{\bibinfo{person}{Werner Rothengatter}.}
  \bibinfo{year}{2007}\natexlab{}.
\newblock \showarticletitle{{Assessment of the impact of renewable electricity
  generation on the German electricity sector An agent-based simulation
  approach}}.
\newblock  (\bibinfo{year}{2007}).
\newblock


\bibitem[\protect\citeauthoryear{Saxena and Abhyankar}{Saxena and
  Abhyankar}{2019}]%
        {Saxena2019}
\bibfield{author}{\bibinfo{person}{Kritika Saxena} {and}
  \bibinfo{person}{Abhijit~R. Abhyankar}.} \bibinfo{year}{2019}\natexlab{}.
\newblock \showarticletitle{{Agent based bilateral transactive market for
  emerging distribution system considering imbalances}}.
\newblock \bibinfo{journal}{\emph{Sustainable Energy, Grids and Networks}}
  \bibinfo{volume}{18} (\bibinfo{year}{2019}), \bibinfo{pages}{100203}.
\newblock
\showISSN{23524677}


\bibitem[\protect\citeauthoryear{Schrattenholzer}{Schrattenholzer}{1981}]%
        {Schrattenholzer1981}
\bibfield{author}{\bibinfo{person}{Leo Schrattenholzer}.}
  \bibinfo{year}{1981}\natexlab{}.
\newblock \showarticletitle{{The energy supply model MESSAGE}}.
\newblock \bibinfo{journal}{\emph{European Journal of Operational Research}}
  \bibinfo{number}{December} (\bibinfo{year}{1981}).
\newblock
\showISBNx{3-704540244}
\showISSN{03772217}


\bibitem[\protect\citeauthoryear{Sun and Tesfatsion}{Sun and
  Tesfatsion}{2007}]%
        {Sun2007}
\bibfield{author}{\bibinfo{person}{Junjie Sun} {and} \bibinfo{person}{Leigh
  Tesfatsion}.} \bibinfo{year}{2007}\natexlab{}.
\newblock \showarticletitle{{Dynamic Testing of Wholesale Power Market Designs
  : An Open-Source Agent-Based Framework}}.
\newblock \bibinfo{journal}{\emph{Computational Economics}}
  \bibinfo{volume}{30}, \bibinfo{number}{3} (\bibinfo{year}{2007}),
  \bibinfo{pages}{291--327}.
\newblock


\bibitem[\protect\citeauthoryear{Weidlich and Veit}{Weidlich and Veit}{2008}]%
        {Weidlich2008}
\bibfield{author}{\bibinfo{person}{Anke Weidlich} {and} \bibinfo{person}{Daniel
  Veit}.} \bibinfo{year}{2008}\natexlab{}.
\newblock \showarticletitle{{A critical survey of agent-based wholesale
  electricity market models}}.
\newblock \bibinfo{journal}{\emph{Energy Economics}} \bibinfo{volume}{30},
  \bibinfo{number}{4} (\bibinfo{year}{2008}), \bibinfo{pages}{1728--1759}.
\newblock
\showISBNx{0140-9883}
\showISSN{01409883}


\bibitem[\protect\citeauthoryear{Wiener}{Wiener}{1930}]%
        {ARIMA}
\bibfield{author}{\bibinfo{person}{N. Wiener}.}
  \bibinfo{year}{1930}\natexlab{}.
\newblock \showarticletitle{Autoregressive integrated moving average}.
\newblock  (\bibinfo{year}{1930}).
\newblock


\bibitem[\protect\citeauthoryear{Zhou, Chan, and Chow}{Zhou
  et~al\mbox{.}}{2007}]%
        {Zhou2007}
\bibfield{author}{\bibinfo{person}{Zhi Zhou}, \bibinfo{person}{Wai~Kin Chan},
  {and} \bibinfo{person}{Joe~H. Chow}.} \bibinfo{year}{2007}\natexlab{}.
\newblock \showarticletitle{{Agent-based simulation of electricity markets: A
  survey of tools}}.
\newblock \bibinfo{journal}{\emph{Artificial Intelligence Review}}
  \bibinfo{volume}{28}, \bibinfo{number}{4} (\bibinfo{year}{2007}),
  \bibinfo{pages}{305--342}.
\newblock
\showISBNx{0269-2821, 1573-7462}
\showISSN{02692821}


\end{thebibliography}

%
\appendix

\section{Research Methods}

Table \ref{table:modern_plant_costs} shows a sample of modern power plant costs, and Table \ref{table:historic_plant_costs} displays a sample of historic power plant costs. The parameters for both of these tables are explained in Section \ref{ssssec:powerplantparameters}

Table \ref{table:scenario_statistics} displays summary statistics for each scenario run. It demonstrates the demand and whether it increases or decreases and by the percentage of change. Carbon tax price is in \textsterling\ per tonne of \ce{CO2}, and also the year range in which the summary statistics apply. 

We then split the low carbon and traditional generation into two groups. Traditional generation contains gas, coal and nuclear power plants, whereas the low carbon group contains photovoltaic as well as offshore and onshore wind turbines. "mean" stands for the arithmetic mean, "std" stands for standard deviation, and min and max are the minimum and maximum values respectively.

\begin{table*}[]
	\begin{tabularx}{1.0205\linewidth}{|l|l|c|l|l|l|l|l|l|l|l|l|l|l|}
	\hline
	Type & Capacity & Year & $\eta$ & $OP$ & $P_D$ & $C_D$ & $P_C$ & $C_C$ & $I_C$ & $F_C$ & $V_C$ & $In_C$ & $Con_C$ \\ \hline
	\multirow{3}{*}{CCGT} & 168.0 & 2018/20/25 & 0.34 & 25 & 3 & 3 & 60,000 & 700,000 & 13,600 & 28,200 & 5 & 2,900 & 3,300 \\ \cline{2-14} 
	& 1200.0 & 2018/20/25 & 0.54 & 25 & 3 & 3 & 10,000 & 500,000 & 15,100 & 12,200 & 3 & 2,100 & 3,300 \\ \cline{2-14} 
	& 1471.0 & 2018/20/25 & 0.53 & 25 & 3 & 3 & 10,000 & 500,000 & 15,100 & 11,400 & 3 & 1,900 & 3,300 \\ \hline
	\multirow{5}{*}{Coal} & 552.0 & 2025 & 0.32 & 25 & 6 & 6 & 40,000 & 3,400,000 & 10,000 & 68,200 & 6 & 13,000 & 3,800 \\ \cline{2-14} 
	& 624.0 & 2025 & 0.32 & 25 & 5 & 5 & 70,000 & 4,200,000 & 10,000 & 79,600 & 3 & 19,300 & 3,800 \\ \cline{2-14} 
	& 652.0 & 2025 & 0.3 & 25 & 5 & 5 & 60,000 & 3,900,000 & 10,000 & 65,300 & 5 & 22,700 & 3,800 \\ \cline{2-14} 
	& 734.0 & 2025 & 0.38 & 25 & 5 & 5 & 60,000 & 2,600,000 & 10,000 & 56,400 & 3 & 9,600 & 3,800 \\ \cline{2-14} 
	& 760.0 & 2025 & 0.35 & 25 & 5 & 5 & 40,000 & 2,800,000 & 10,000 & 52,100 & 5 & 14,000 & 3,800 \\ \hline
	\multirow{3}{*}{Hydro} & 0.033 & 2018/20/25 & 1.0 & 35 & 0 & 0 & 0 & 6,300,000 & 0 & 83,300 & 0 & 0 & 0 \\ \cline{2-14} 
	& 1.046 & 2018/20/25 & 1.0 & 35 & 0 & 0 & 0 & 3,300,000 & 400 & 18,200 & 0 & 0 & 0 \\ \cline{2-14} 
	& 11.0 & 2018/20/25 & 1.0 & 41 & 2 & 2 & 60,000 & 3,000,000 & 0 & 45,100 & 6 & 0 & 0 \\ \hline
	Nuclear & 3300.0 & 2025 & 1.0 & 60 & 5 & 8 & 240,000 & 4,100,000 & 11,500 & 72,900 & 5 & 10,000 & 500 \\ \hline
	\multirow{5}{*}{OCGT} & 96.0 & 2018/20/25 & 0.35 & 25 & 2 & 2 & 80,000 & 600,000 & 12,600 & 9,900 & 4 & 2,500 & 2,400 \\ \cline{2-14} 
	& 299.0 & 2018/20/25 & 0.35 & 25 & 2 & 2 & 30,000 & 400,000 & 13,600 & 9,600 & 3 & 1,600 & 2,500 \\ \cline{2-14} 
	& 311.0 & 2018/20/25 & 0.35 & 25 & 2 & 2 & 30,000 & 400,000 & 13,600 & 9,500 & 3 & 1,600 & 2,500 \\ \cline{2-14} 
	& 400.0 & 2018/20/25 & 0.34 & 25 & 2 & 2 & 30,000 & 300,000 & 15,100 & 7,800 & 3 & 1,300 & 2,500 \\ \cline{2-14} 
	& 625.0 & 2018/20/25 & 0.35 & 25 & 2 & 2 & 20,000 & 300,000 & 15,100 & 4,600 & 3 & 1,200 & 2,400 \\ \hline
	\multirow{6}{*}{Offshore} & \multirow{3}{*}{321.0} & 2018 & 0.0 & 23 & 5 & 3 & 60,000 & 2,200,000 & 69,300 & 30,900 & 3 & 1,400 & 33,500 \\ \cline{3-14} 
	&  & 2020 & 0.0 & 23 & 5 & 3 & 60,000 & 2,100,000 & 69,300 & 30,000 & 3 & 1,400 & 32,600 \\ \cline{3-14} 
	&  & 2025 & 0.0 & 23 & 5 & 3 & 60,000 & 1,900,000 & 69,300 & 28,600 & 3 & 1,300 & 31,100 \\ \cline{2-14} 
	& \multirow{3}{*}{844.0} & 2018 & 0.0 & 22 & 5 & 3 & 120,000 & 2,400,000 & 323,000 & 48,600 & 4 & 3,300 & 50,300 \\ \cline{3-14} 
	&  & 2020 & 0.0 & 22 & 5 & 3 & 120,000 & 2,300,000 & 323,000 & 47,300 & 3 & 3,300 & 48,900 \\ \cline{3-14} 
	&  & 2025 & 0.0 & 22 & 5 & 3 & 120,000 & 2,100,000 & 323,000 & 45,400 & 3 & 3,100 & 47,000 \\ \hline
	\multirow{9}{*}{Onshore} & \multirow{3}{*}{0.01} & 2018 & 1.0 & 20 & 0 & 0 & 0 & 3,700,000 & 0 & 29,700 & 0 & 0 & 0 \\ \cline{3-14} 
	&  & 2020 & 1.0 & 20 & 0 & 0 & 0 & 3,600,000 & 0 & 29,600 & 0 & 0 & 0 \\ \cline{3-14} 
	&  & 2025 & 1.0 & 20 & 0 & 0 & 0 & 3,500,000 & 0 & 29,600 & 0 & 0 & 0 \\ \cline{2-14} 
	& \multirow{3}{*}{0.482} & 2018 & 1.0 & 20 & 0 & 0 & 0 & 2,200,000 & 200 & 56,900 & 0 & 0 & 0 \\ \cline{3-14} 
	&  & 2020 & 1.0 & 20 & 0 & 0 & 0 & 2,100,000 & 200 & 56,900 & 0 & 0 & 0 \\ \cline{3-14} 
	&  & 2025 & 1.0 & 20 & 0 & 0 & 0 & 2,000,000 & 200 & 56,700 & 0 & 0 & 0 \\ \cline{2-14} 
	& \multirow{3}{*}{20.0} & 2018 & 0.0 & 24 & 4 & 2 & 110,000 & 1,200,000 & 3,300 & 23,200 & 5 & 1,400 & 3,100 \\ \cline{3-14} 
	&  & 2020 & 0.0 & 24 & 4 & 2 & 110,000 & 1,200,000 & 3,300 & 23,000 & 5 & 1,400 & 3,100 \\ \cline{3-14} 
	&  & 2025 & 0.0 & 24 & 4 & 2 & 110,000 & 1,200,000 & 3,300 & 22,400 & 5 & 1,400 & 3,000 \\ \hline
	\multirow{14}{*}{PV} & \multirow{3}{*}{0.003} & 2018 & 1.0 & 30 & 0 & 0 & 0 & 1,500,000 & 0 & 23,500 & 0 & 0 & 0 \\ \cline{3-14} 
	&  & 2020 & 1.0 & 30 & 0 & 0 & 0 & 1,500,000 & 0 & 23,400 & 0 & 0 & 0 \\ \cline{3-14} 
	&  & 2025 & 1.0 & 30 & 0 & 0 & 0 & 1,400,000 & 0 & 23,200 & 0 & 0 & 0 \\ \cline{2-14} 
	& \multirow{2}{*}{0.455} & 2018 & 1.0 & 30 & 0 & 0 & 0 & 1,000,000 & 200 & 9,400 & 0 & 0 & 0 \\ \cline{3-14} 
	&  & 2025 & 1.0 & 30 & 0 & 0 & 0 & 900,000 & 200 & 9,200 & 0 & 0 & 0 \\ \cline{2-14} 
	& \multirow{3}{*}{1.0} & 2018 & 0.0 & 25 & 1 & 0 & 20,000 & 700,000 & 0 & 6,600 & 3 & 2,600 & 1,300 \\ \cline{3-14} 
	&  & 2020 & 0.0 & 25 & 1 & 0 & 20,000 & 700,000 & 0 & 6,300 & 3 & 2,600 & 1,300 \\ \cline{3-14} 
	&  & 2025 & 0.0 & 25 & 1 & 0 & 20,000 & 600,000 & 0 & 5,900 & 3 & 2,400 & 1,200 \\ \cline{2-14} 
	& \multirow{3}{*}{4.0} & 2018 & 0.0 & 25 & 1 & 0 & 60,000 & 700,000 & 200 & 8,300 & 0 & 1,200 & 1,300 \\ \cline{3-14} 
	&  & 2020 & 0.0 & 25 & 1 & 0 & 60,000 & 700,000 & 200 & 8,000 & 0 & 1,100 & 1,300 \\ \cline{3-14} 
	&  & 2025 & 0.0 & 25 & 1 & 0 & 60,000 & 600,000 & 200 & 7,500 & 0 & 1,100 & 1,200 \\ \cline{2-14} 
	& \multirow{3}{*}{16.0} & 2018 & 0.0 & 25 & 1 & 0 & 70,000 & 700,000 & 400 & 5,600 & 0 & 2,000 & 1,300 \\ \cline{3-14} 
	&  & 2020 & 0.0 & 25 & 1 & 0 & 70,000 & 600,000 & 400 & 5,400 & 0 & 1,900 & 1,300 \\ \cline{3-14} 
	&  & 2025 & 0.0 & 25 & 1 & 0 & 70,000 & 600,000 & 400 & 5,100 & 0 & 1,800 & 1,200 \\ \hline
	Recip. Engine (Diesel) & 20.0 & 2018/20/25 & 0.34 & 15 & 2 & 1 & 10,000 & 300,000 & 2,200 & 10,000 & 2 & 1,000 & -31,900 \\ \hline
	Recip. Engine (Gas) & 20.0 & 2018/20/25 & 0.32 & 15 & 2 & 1 & 10,000 & 300,000 & 3,400 & 10,000 & 2 & 1,000 & -31,900 \\ \hline
		
	\end{tabularx}

	\caption{Modern power plant costs \cite{Department2016}}
	\label{table:modern_plant_costs}
\end{table*}

\begin{table*}[]
	\begin{tabular}{|l|l|l|l|l|l|l|l|l|l|l|l|l|l|}
	\hline
	Type & Capacity & Year & $\eta$ & $OP$ & $P_D$ & $C_D$ & $P_C$ & $C_C$ & $I_C$ & $F_C$ & $V_C$ & $In_C$ & $Con_C$ \\ \hline
	\multirow{12}{*}{CCGT} & \multirow{4}{*}{168.0} & 1980 & 0.34 & 25 & 3 & 3 & 207,345 & 2,419,027 & 46,998 & 97,452 & 22 & 10,021 & 11,403 \\ \cline{3-14} 
	&  & 1990 & 0.34 & 25 & 3 & 3 & 181,208 & 2,114,099 & 41,073 & 85,167 & 13 & 8,758 & 9,966 \\ \cline{3-14} 
	&  & 2000 & 0.34 & 25 & 3 & 3 & 116,407 & 1,358,089 & 26,385 & 54,711 & 10 & 5,626 & 6,402 \\ \cline{3-14} 
	&  & 2010 & 0.34 & 25 & 3 & 3 & 73,530 & 857,857 & 16,666 & 34,559 & 11 & 3,553 & 4,044 \\ \cline{2-14} 
	& \multirow{4}{*}{1200.0} & 1980 & 0.54 & 25 & 3 & 3 & 59,102 & 2,955,138 & 89,245 & 72,105 & 31 & 12,411 & 19,503 \\ \cline{3-14} 
	&  & 1990 & 0.54 & 25 & 3 & 3 & 59,884 & 2,994,246 & 90,426 & 73,059 & 21 & 12,575 & 19,762 \\ \cline{3-14} 
	&  & 2000 & 0.54 & 25 & 3 & 3 & 49,674 & 2,483,747 & 75,009 & 60,603 & 21 & 10,431 & 16,392 \\ \cline{3-14} 
	&  & 2010 & 0.54 & 25 & 3 & 3 & 60,640 & 3,032,008 & 91,566 & 73,981 & 13 & 12,734 & 20,011 \\ \cline{2-14} 
	& \multirow{4}{*}{1471.0} & 1980 & 0.53 & 25 & 3 & 3 & 92,000 & 4,600,023 & 138,920 & 104,880 & 10 & 17,480 & 30,360 \\ \cline{3-14} 
	&  & 1990 & 0.53 & 25 & 3 & 3 & 54,296 & 2,714,817 & 81,987 & 61,897 & 26 & 10,316 & 17,917 \\ \cline{3-14} 
	&  & 2000 & 0.53 & 25 & 3 & 3 & 49,310 & 2,465,515 & 74,458 & 56,213 & 21 & 9,368 & 16,272 \\ \cline{3-14} 
	&  & 2010 & 0.53 & 25 & 3 & 3 & 46,998 & 2,349,947 & 70,968 & 53,578 & 21 & 8,929 & 15,509 \\ \hline
	\multirow{24}{*}{Coal} & \multirow{4}{*}{552.0} & 1980 & 0.32 & 25 & 6 & 6 & 118,041 & 10,033,488 & 29,510 & 201,259 & 22 & 38,363 & 11,213 \\ \cline{3-14} 
	&  & 1990 & 0.32 & 25 & 6 & 6 & 41,766 & 3,550,192 & 10,441 & 71,212 & 2 & 13,574 & 3,967 \\ \cline{3-14} 
	&  & 2000 & 0.32 & 25 & 6 & 6 & 51,429 & 4,371,538 & 12,857 & 87,687 & 3 & 16,714 & 4,885 \\ \cline{3-14} 
	&  & 2010 & 0.32 & 25 & 6 & 6 & 43,411 & 3,689,957 & 10,852 & 74,016 & 10 & 14,108 & 4,124 \\ \cline{2-14} 
	& \multirow{8}{*}{624.0} & 1980 & 0.32 & 25 & 5 & 5 & 183,851 & 11,031,076 & 26,264 & 206,176 & 15 & 41,497 & 9,980 \\ \cline{3-14} 
	&  & 1980 & 0.32 & 25 & 5 & 5 & 188,476 & 11,308,571 & 26,925 & 211,362 & 11 & 42,541 & 10,231 \\ \cline{3-14} 
	&  & 1990 & 0.32 & 25 & 5 & 5 & 62,458 & 3,747,483 & 8,922 & 70,042 & 5 & 14,097 & 3,390 \\ \cline{3-14} 
	&  & 1990 & 0.32 & 25 & 5 & 5 & 65,126 & 3,907,588 & 9,303 & 73,034 & 3 & 14,699 & 3,535 \\ \cline{3-14} 
	&  & 2000 & 0.32 & 25 & 5 & 5 & 80,033 & 4,802,002 & 11,433 & 89,751 & 3 & 18,064 & 4,344 \\ \cline{3-14} 
	&  & 2000 & 0.32 & 25 & 5 & 5 & 80,882 & 4,852,979 & 11,554 & 90,704 & 3 & 18,256 & 4,390 \\ \cline{3-14} 
	&  & 2010 & 0.32 & 25 & 5 & 5 & 84,549 & 5,072,973 & 12,078 & 94,816 & 3 & 19,084 & 4,589 \\ \cline{3-14} 
	&  & 2010 & 0.32 & 25 & 5 & 5 & 81,834 & 4,910,056 & 11,690 & 91,771 & 5 & 18,471 & 4,442 \\ \cline{2-14} 
	& \multirow{4}{*}{652.0} & 1980 & 0.3 & 25 & 5 & 5 & 161,344 & 10,487,387 & 26,890 & 175,596 & 16 & 61,041 & 10,218 \\ \cline{3-14} 
	&  & 1990 & 0.3 & 25 & 5 & 5 & 54,542 & 3,545,235 & 9,090 & 59,359 & 4 & 20,635 & 3,454 \\ \cline{3-14} 
	&  & 2000 & 0.3 & 25 & 5 & 5 & 68,516 & 4,453,581 & 11,419 & 74,568 & 2 & 25,922 & 4,339 \\ \cline{3-14} 
	&  & 2010 & 0.3 & 25 & 5 & 5 & 67,915 & 4,414,497 & 11,319 & 73,914 & 4 & 25,694 & 4,301 \\ \cline{2-14} 
	& \multirow{4}{*}{734.0} & 1980 & 0.38 & 25 & 5 & 5 & 249,766 & 10,823,198 & 41,627 & 234,780 & 16 & 39,962 & 15,818 \\ \cline{3-14} 
	&  & 1990 & 0.38 & 25 & 5 & 5 & 87,920 & 3,809,903 & 14,653 & 82,645 & 7 & 14,067 & 5,568 \\ \cline{3-14} 
	&  & 2000 & 0.38 & 25 & 5 & 5 & 118,072 & 5,116,482 & 19,678 & 110,988 & 5 & 18,891 & 7,477 \\ \cline{3-14} 
	&  & 2010 & 0.38 & 25 & 5 & 5 & 132,370 & 5,736,075 & 22,061 & 124,428 & 5 & 21,179 & 8,383 \\ \cline{2-14} 
	& \multirow{4}{*}{760.0} & 1980 & 0.35 & 25 & 5 & 5 & 160,182 & 11,212,746 & 40,045 & 208,637 & 8 & 56,063 & 15,217 \\ \cline{3-14} 
	&  & 1990 & 0.35 & 25 & 5 & 5 & 55,208 & 3,864,573 & 13,802 & 71,908 & 4 & 19,322 & 5,244 \\ \cline{3-14} 
	&  & 2000 & 0.35 & 25 & 5 & 5 & 65,705 & 4,599,358 & 16,426 & 85,580 & 8 & 22,996 & 6,241 \\ \cline{3-14} 
	&  & 2010 & 0.35 & 25 & 5 & 5 & 77,393 & 5,417,570 & 19,348 & 100,805 & 3 & 27,087 & 7,352 \\ \hline
	\multirow{4}{*}{Nuclear} & \multirow{4}{*}{3300.0} & 1980 & 1.0 & 60 & 5 & 8 & 516,790 & 8,828,507 & 24,762 & 156,975 & 21 & 21,532 & 1,076 \\ \cline{3-14} 
	&  & 1990 & 1.0 & 60 & 5 & 8 & 390,159 & 6,665,224 & 18,695 & 118,510 & 3 & 16,256 & 812 \\ \cline{3-14} 
	&  & 2000 & 1.0 & 60 & 5 & 8 & 378,998 & 6,474,560 & 18,160 & 115,120 & 15 & 15,791 & 789 \\ \cline{3-14} 
	&  & 2010 & 1.0 & 60 & 5 & 8 & 388,457 & 6,636,156 & 18,613 & 117,994 & 13 & 16,185 & 809 \\ \hline
	\multirow{8}{*}{Offshore} & \multirow{4}{*}{321.0} & 1980 & 0.0 & 23 & 5 & 3 & 100,043 & 3,668,254 & 115,550 & 51,522 & 9 & 2,334 & 55,857 \\ \cline{3-14} 
	&  & 1990 & 0.0 & 23 & 5 & 3 & 104,550 & 3,833,513 & 120,755 & 53,843 & 3 & 2,439 & 58,373 \\ \cline{3-14} 
	&  & 2000 & 0.0 & 23 & 5 & 3 & 102,374 & 3,753,742 & 118,242 & 52,723 & 6 & 2,388 & 57,159 \\ \cline{3-14} 
	&  & 2010 & 0.0 & 23 & 5 & 3 & 98,571 & 3,614,292 & 113,850 & 50,764 & 6 & 2,300 & 55,035 \\ \cline{2-14} 
	& \multirow{4}{*}{844.0} & 1980 & 0.0 & 22 & 5 & 3 & 181,469 & 3,629,393 & 488,455 & 73,495 & 8 & 4,990 & 76,066 \\ \cline{3-14} 
	&  & 1990 & 0.0 & 22 & 5 & 3 & 178,822 & 3,576,447 & 481,330 & 72,423 & 10 & 4,917 & 74,956 \\ \cline{3-14} 
	&  & 2000 & 0.0 & 22 & 5 & 3 & 180,212 & 3,604,250 & 485,072 & 72,986 & 9 & 4,955 & 75,539 \\ \cline{3-14} 
	&  & 2010 & 0.0 & 22 & 5 & 3 & 171,372 & 3,427,446 & 461,277 & 69,405 & 11 & 4,712 & 71,833 \\ \hline
	\multirow{4}{*}{Onshore} & \multirow{4}{*}{20.0} & 1980 & 0.0 & 24 & 4 & 2 & 374,087 & 4,080,950 & 11,222 & 78,898 & 26 & 4,761 & 10,542 \\ \cline{3-14} 
	&  & 1990 & 0.0 & 24 & 4 & 2 & 411,234 & 4,486,197 & 12,337 & 86,733 & 10 & 5,233 & 11,589 \\ \cline{3-14} 
	&  & 2000 & 0.0 & 24 & 4 & 2 & 230,491 & 2,514,457 & 6,914 & 48,612 & 5 & 2,933 & 6,495 \\ \cline{3-14} 
	&  & 2010 & 0.0 & 24 & 4 & 2 & 143,450 & 1,564,915 & 4,303 & 30,255 & 7 & 1,825 & 4,042 \\ \hline
	\multirow{4}{*}{PV} & \multirow{4}{*}{16.0} & 1980 & 0.0 & 25 & 1 & 0 & 399,799 & 3,997,991 & 2,284 & 31,983 & 0 & 11,422 & 7,424 \\ \cline{3-14} 
	&  & 1990 & 0.0 & 25 & 1 & 0 & 399,799 & 3,997,991 & 2,284 & 31,983 & 0 & 11,422 & 7,424 \\ \cline{3-14} 
	&  & 2000 & 0.0 & 25 & 1 & 0 & 399,799 & 3,997,991 & 2,284 & 31,983 & 0 & 11,422 & 7,424 \\ \cline{3-14} 
	&  & 2010 & 0.0 & 25 & 1 & 0 & 399,799 & 3,997,991 & 2,284 & 31,983 & 0 & 11,422 & 7,424 \\ \hline
	\end{tabular}
	\caption{Sample of historic power plant costs \cite{IRENA2018,IEA2015,OWPB2016}} 
	\label{table:historic_plant_costs}

\end{table*}

\clearpage



\begin{table}[H]
	\begin{tabular}{|l|l|l|l|l|l|l|l|l|l|l|}
		\hline
		\multirow{2}{*}{\textbf{Demand}} & \multirow{2}{*}{\textbf{Carbon Tax}} & \multirow{2}{*}{\textbf{Year Range}} & \multicolumn{4}{l|}{\textbf{Low Carbon}} & \multicolumn{4}{l|}{\textbf{Traditional Generation}} \\ \cline{4-11} 
		&  &  & \textbf{mean} & \textbf{std} & \textbf{min} & \textbf{max} & \textbf{mean} & \textbf{std} & \textbf{min} & \textbf{max} \\ \hline
		\multirow{24}{*}{Demand Decreasing 1\% a Year} & \multirow{3}{*}{0} & 2019-2029 & 14.14 & 5.16 & 6.36 & 27.29 & 85.86 & 5.16 & 72.71 & 93.64 \\ \cline{3-11} 
		&  & 2029-2039 & 16.95 & 11.19 & 6.2 & 52.52 & 83.05 & 11.19 & 47.48 & 93.8 \\ \cline{3-11} 
		&  & 2039-2050 & 22.29 & 18.01 & 4.72 & 60.0 & 77.71 & 18.01 & 40.0 & 95.28 \\ \cline{2-11} 
		& \multirow{3}{*}{10} & 2019-2029 & 15.85 & 8.82 & 8.8 & 41.0 & 84.15 & 8.82 & 59.0 & 91.2 \\ \cline{3-11} 
		&  & 2029-2039 & 20.33 & 15.34 & 7.92 & 62.75 & 79.67 & 15.34 & 37.25 & 92.08 \\ \cline{3-11} 
		&  & 2039-2050 & 24.38 & 17.17 & 8.79 & 61.87 & 75.62 & 17.17 & 38.13 & 91.21 \\ \cline{2-11} 
		& \multirow{3}{*}{170 to 22} & 2019-2029 & 92.03 & 8.32 & 71.2 & 99.8 & 7.97 & 8.32 & 0.2 & 28.8 \\ \cline{3-11} 
		&  & 2029-2039 & 99.66 & 0.11 & 99.11 & 99.82 & 0.34 & 0.11 & 0.18 & 0.89 \\ \cline{3-11} 
		&  & 2039-2050 & 99.59 & 0.1 & 99.32 & 99.75 & 0.41 & 0.1 & 0.25 & 0.68 \\ \cline{2-11} 
		& \multirow{3}{*}{26 to 174} & 2019-2029 & 24.84 & 11.32 & 11.01 & 65.78 & 75.16 & 11.32 & 34.22 & 88.99 \\ \cline{3-11} 
		&  & 2029-2039 & 42.6 & 21.63 & 11.28 & 79.05 & 57.4 & 21.63 & 20.95 & 88.72 \\ \cline{3-11} 
		&  & 2039-2050 & 56.42 & 15.48 & 31.63 & 81.72 & 43.58 & 15.48 & 18.28 & 68.37 \\ \cline{2-11} 
		& \multirow{3}{*}{20} & 2019-2029 & 22.94 & 11.92 & 7.8 & 62.07 & 77.06 & 11.92 & 37.93 & 92.2 \\ \cline{3-11} 
		&  & 2029-2039 & 40.52 & 21.73 & 7.04 & 73.0 & 59.48 & 21.73 & 27.0 & 92.96 \\ \cline{3-11} 
		&  & 2039-2050 & 49.36 & 20.73 & 10.82 & 79.09 & 50.64 & 20.73 & 20.91 & 89.18 \\ \cline{2-11} 
		& \multirow{3}{*}{40} & 2019-2029 & 48.16 & 12.28 & 32.61 & 82.35 & 51.84 & 12.28 & 17.65 & 67.39 \\ \cline{3-11} 
		&  & 2029-2039 & 69.08 & 12.12 & 46.05 & 93.13 & 30.92 & 12.12 & 6.87 & 53.95 \\ \cline{3-11} 
		&  & 2039-2050 & 70.61 & 10.82 & 52.5 & 91.98 & 29.39 & 10.82 & 8.02 & 47.5 \\ \cline{2-11} 
		& \multirow{3}{*}{50} & 2019-2029 & 53.78 & 23.42 & 17.98 & 92.93 & 46.22 & 23.42 & 7.07 & 82.02 \\ \cline{3-11} 
		&  & 2029-2039 & 68.41 & 20.18 & 29.54 & 96.29 & 31.59 & 20.18 & 3.71 & 70.46 \\ \cline{3-11} 
		&  & 2039-2050 & 66.86 & 20.42 & 38.31 & 99.73 & 33.14 & 20.42 & 0.27 & 61.69 \\ \cline{2-11} 
		& \multirow{3}{*}{70} & 2019-2029 & 83.62 & 13.16 & 41.29 & 99.41 & 16.38 & 13.16 & 0.59 & 58.71 \\ \cline{3-11} 
		&  & 2029-2039 & 96.76 & 4.43 & 83.93 & 99.99 & 3.24 & 4.43 & 0.01 & 16.07 \\ \cline{3-11} 
		&  & 2039-2050 & 97.63 & 3.58 & 87.8 & 99.94 & 2.37 & 3.58 & 0.06 & 12.2 \\ \hline
		\multirow{24}{*}{Demand Increasing 1\% a Year} & \multirow{3}{*}{0} & 2019-2029 & 14.87 & 9.9 & 6.73 & 45.59 & 85.13 & 9.9 & 54.41 & 93.27 \\ \cline{3-11} 
		&  & 2029-2039 & 17.07 & 16.39 & 4.8 & 65.87 & 82.93 & 16.39 & 34.13 & 95.2 \\ \cline{3-11} 
		&  & 2039-2050 & 17.54 & 20.0 & 3.83 & 67.95 & 82.46 & 20.0 & 32.05 & 96.17 \\ \cline{2-11} 
		& \multirow{3}{*}{10} & 2019-2029 & 18.96 & 7.17 & 10.23 & 39.02 & 81.04 & 7.17 & 60.98 & 89.77 \\ \cline{3-11} 
		&  & 2029-2039 & 23.44 & 16.47 & 8.89 & 61.96 & 76.56 & 16.47 & 38.04 & 91.11 \\ \cline{3-11} 
		&  & 2039-2050 & 27.91 & 19.45 & 9.64 & 67.06 & 72.09 & 19.45 & 32.94 & 90.36 \\ \cline{2-11} 
		& \multirow{3}{*}{170 to 22} & 2019-2029 & 92.09 & 9.29 & 67.32 & 99.8 & 7.91 & 9.29 & 0.2 & 32.68 \\ \cline{3-11} 
		&  & 2029-2039 & 99.98 & 0.05 & 99.76 & 100.0 & 0.02 & 0.05 & 0.0 & 0.24 \\ \cline{3-11} 
		&  & 2039-2050 & 100.0 & 0.0 & 100.0 & 100.0 & 0.0 & 0.0 & 0.0 & 0.0 \\ \cline{2-11} 
		& \multirow{3}{*}{26 to 174} & 2019-2029 & 24.75 & 11.33 & 11.95 & 56.65 & 75.25 & 11.33 & 43.35 & 88.05 \\ \cline{3-11} 
		&  & 2029-2039 & 39.28 & 20.39 & 10.87 & 73.41 & 60.72 & 20.39 & 26.59 & 89.13 \\ \cline{3-11} 
		&  & 2039-2050 & 49.72 & 18.84 & 22.02 & 86.43 & 50.28 & 18.84 & 13.57 & 77.98 \\ \cline{2-11} 
		& \multirow{3}{*}{20} & 2019-2029 & 26.32 & 16.01 & 8.08 & 83.77 & 73.68 & 16.01 & 16.23 & 91.92 \\ \cline{3-11} 
		&  & 2029-2039 & 37.21 & 23.72 & 5.2 & 82.72 & 62.79 & 23.72 & 17.28 & 94.8 \\ \cline{3-11} 
		&  & 2039-2050 & 45.79 & 26.31 & 7.5 & 88.24 & 54.21 & 26.31 & 11.76 & 92.5 \\ \cline{2-11} 
		& \multirow{3}{*}{40} & 2019-2029 & 43.41 & 18.58 & 13.96 & 80.7 & 56.59 & 18.58 & 19.3 & 86.04 \\ \cline{3-11} 
		&  & 2029-2039 & 61.79 & 29.18 & 14.83 & 92.44 & 38.21 & 29.18 & 7.56 & 85.17 \\ \cline{3-11} 
		&  & 2039-2050 & 75.03 & 23.95 & 21.4 & 95.91 & 24.97 & 23.95 & 4.09 & 78.6 \\ \cline{2-11} 
		& \multirow{3}{*}{50} & 2019-2029 & 64.64 & 23.56 & 16.96 & 99.22 & 35.36 & 23.56 & 0.78 & 83.04 \\ \cline{3-11} 
		&  & 2029-2039 & 86.48 & 16.8 & 23.27 & 99.44 & 13.52 & 16.8 & 0.56 & 76.73 \\ \cline{3-11} 
		&  & 2039-2050 & 91.18 & 9.17 & 65.77 & 99.78 & 8.82 & 9.17 & 0.22 & 34.23 \\ \cline{2-11} 
		& \multirow{3}{*}{70} & 2019-2029 & 69.61 & 19.77 & 26.36 & 100.0 & 30.39 & 19.77 & 0.0 & 73.64 \\ \cline{3-11} 
		&  & 2029-2039 & 89.07 & 13.79 & 31.57 & 100.0 & 10.93 & 13.79 & 0.0 & 68.43 \\ \cline{3-11} 
		&  & 2039-2050 & 91.77 & 10.37 & 67.5 & 100.0 & 8.23 & 10.37 & 0.0 & 32.5 \\ \hline
	\end{tabular}
	\caption{Summary statistics for each scenario run.}
	
	\label{table:scenario_statistics} 
	
\end{table}

\end{document}